\documentclass[]{aastex7}
\usepackage{subfig}
\usepackage{rotating}
\usepackage{array}
\usepackage{longtable}
\usepackage{booktabs}
\usepackage{threeparttable}
\usepackage{appendix}
\usepackage{float}
\usepackage{listings}
\begin{document}
\title{Demystifying image-recovery from radio interferometers: toward a multiscale predictive model}
\correspondingauthor{Guang-Xing Li}
\email{gxli@ynu.edu.cn}
\author[0000-0001-9798-9852]{Dan Miao}
\affiliation{Department of Astronomy, Xiamen University, Zengcuo’an West Road, Xiamen, 361005, China}
\email{miaodan@stu.xmu.edu.cn}
\author[0000-0003-3144-1952]{Guang-Xing Li}
\affiliation{South-Western Institute for Astronomy Research, Yunnan University, Chenggong District, Kunming 650500, China}
\email{gxli@ynu.edu.cn}
\begin{abstract}
Radio interferometers suffer from the missing short-spacing problem, losing large-scale diffuse emission.
This missing flux underestimates gas mass and biases key metrics like star formation efficiency.
Quantifying this scale-dependent loss currently relies on computationally intensive mock observations, lacking an analytical image-domain framework.
We introduce the Constrained Diffusion Decomposition (CDD) method to decompose an input image ($I_{\mathrm{in}}$) into $n$ continuous scale-space components, denoted as $I_l = \mathrm{CDD}_l(I_{\mathrm{in}})$ for $l \in [1, n]$, and apply it to simulated Atacama Large Millimeter/submillimeter Array (ALMA) observations of the Perseus molecular cloud across multiple array configurations.
We find that the interferometric spatial filtering response can be mathematically decoupled: the scale-dependent flux recovery fraction follows a one-dimensional error function (\texttt{erf}), defined as $R(l) = \frac{B}{2} \left[ 1 - \mathrm{erf}\left( \frac{l - c_{\mathrm{recover}}}{w} \right) \right]$, where compact structures are effectively recovered, while extended emission decays monotonically as scales approach the maximum recoverable scale.
The proposed CDD--\texttt{erf} framework predicts the spatially filtered interferometric image $I_{\mathrm{pred}}$ directly in the image domain, bypassing visibility simulations, mapping the true sky brightness distribution via the equation $I_{\mathrm{pred}} = \sum_{l=1}^{n} [ \mathrm{CDD}_l(I_{\mathrm{in}}) \times R(l)]$.
This provides a quantitative bridge between model and interferometric observations.
\end{abstract}

\section{Introduction} \label{sec:intro} \setcounter{footnote}{0}
Radio waves constitute a fundamental component of the electromagnetic spectrum.
In radio astronomy, a substantial fraction of our astrophysical knowledge is derived from radio interferometric observations.
While interferometric data are intrinsically acquired as sparsely sampled visibilities in the Fourier domain, the physical interpretation of the universe relies on analyzing continuous spatial structures in the image domain.
Understanding the correspondence between the image and Fourier domains is therefore essential.
In practice, radio interferometers probe the true sky brightness distribution by sampling visibilities in the Fourier $uv$-plane.
However, reconstructing a continuous image from these discrete measurements is a fundamentally ill-posed mathematical problem.
According to the van~Cittert--Zernike theorem, the sampled spatial scales are inversely proportional to the projected baseline lengths \citep{2017isra.book.....T}.
Consequently, the observable spatial scales are truncated at both ends: the longest baseline sets the angular resolution, while the shortest baseline determines the maximum recoverable scale (MRS, $\theta_{\mathrm{MRS}}$; \citealt{cortes_2025_14933753}).
Furthermore, an array of $N$ antennas can instantaneously measure only $N(N-1)/2$ discrete Fourier components.
As a result, only a fraction of the spatial frequency components is sampled at any given time, making the $uv$-coverage intrinsically sparse.
Radio interferometric arrays therefore function as sparse spatial-frequency bandpass filters.
In particular, the finite physical diameter of individual antennas prevents the sampling of the shortest baselines and the origin of the $uv$-plane, giving rise to the well-known zero-spacing problem \citep{2002ASPC..278..375S}.
Consequently, interferometers are insensitive to large-scale diffuse emission.
When the angular scales of the emission approach or exceed the MRS, the extended structures are resolved out, and only compact features are retained in the observed data.
Image reconstruction aims to recover a continuous brightness map from sparsely sampled Fourier components.
Among the observational limitations inherent to interferometry, the loss of large-scale diffuse emission is particularly consequential.
Extended structures correspond to low spatial frequencies, which require densely packed antenna configurations for adequate sampling.
However, achieving such configurations is physically constrained by effects such as antenna shadowing.
Mathematically, the unsampled gaps in the $uv$-plane are typically set to zero during image reconstruction.
As a result, the dirty image is the convolution of the true sky brightness distribution with the synthesized beam, which introduces sidelobes and imaging artifacts \citep{2017isra.book.....T}.
Deconvolution algorithms, such as \textsc{clean} \citep{1974A&AS...15..417H,1984AJ.....89.1076S,2022PASP..134k4501C}, are routinely applied to mitigate these artifacts; however, they are inherently non-linear and depend on user-defined parameters.
As a consequence, large-scale diffuse structures remain fundamentally unrecoverable in standard interferometric observations.
To avoid the uncertainties associated with missing flux, studies are generally restricted to compact features whose angular scales fall well below the MRS, where flux recovery can be considered reliable.
While simply acknowledging this observational limitation may be acceptable for isolated, point-like sources that fall well within the resolution and sensitivity limits of current interferometric arrays, this approach becomes inadequate when studying complex gas structures such as galaxies or molecular clouds.
These astrophysical systems are inherently multiscale, governed by physical scaling relations such as Larson's laws \citep{1981MNRAS.194..809L}.
Over the years, various data combination techniques have been developed to recover missing flux by incorporating single-dish observations---provided such complementary data are available---such as combination before image deconvolution (e.g., \texttt{SDINT}, \citealt{2019AJ....158....3R}; and \texttt{TP2VIS}, \citealt{2019PASP..131e4505K}), during deconvolution (e.g., \texttt{MACF}, \citealt{2015AJ....150..159D, 2017A&A...603A..89K, 2023PASP..135c4501P}), and after deconvolution (e.g., \texttt{Feather}, \citealt{2017PASP..129i4501C}; and FSSC, \citealt{2018AN....339...87F}) classification by \citet{2023PASP..135c4501P}.
However, when single-dish data are lacking, a quantitative characterization and prediction of the missing flux becomes necessary to accurately interpret the observed structures.

Previously, this has relied on computationally expensive mock observations, such as those performed using the Common Astronomy Software Applications (\textsc{casa}) simulation pipeline\footnote{Simulations of Atacama Large Millimeter/submillimeter Array (ALMA) observations were performed using the \textsc{casa} package:
\url{https://casaguides.nrao.edu/index.php/Simulating_Observations_in_CASA}} \citep{2022PASP..134k4501C}.
As a result, the quantification of flux loss lacks an analytical mathematical foundation in the image domain, making it difficult to characterize flux attenuation as a continuous function of spatial scale.
In this paper, we present a new approach that provides an analytical framework for understanding interferometric flux loss directly in the image domain.
To establish an analytical foundation in the image domain, we introduce the Constrained Diffusion Decomposition \citep[CDD;][]{2022ApJS..259...59L} method\footnote{The CDD Python package is available at \url{https://github.com/gxli/Constrained-Diffusion-Decomposition}.}.
Rather than attempting to disentangle the complex convolution in the image plane directly, CDD projects the observed image into a continuous ``scale-space''.
By solving a modified, nonlinear diffusion equation, CDD decomposes astronomical signals --- including, e.g., one-dimensional spectra, two-dimensional intensity maps, and three-dimensional data cubes --- into a set of independent sub-images, each of which isolates structures at a specific spatial scale.
Many astrophysical processes exhibit multiscale structure, making scale decomposition a necessary step in their analysis.
Although traditional wavelet transforms can separate multiscale components, they inherently produce artifacts in the form of unphysical negative values near regions with sharp intensity gradients \citep{2006aida.book.....S}.
The CDD method avoids this limitation by construction: the decomposed sub-images are guaranteed to retain strictly non-negative values at all scales, ensuring that the decomposition remains physically meaningful for analyzing the distribution of both radiation and matter.
In this work, we formalize this scale-space perspective into the \textbf{CDD-\texttt{erf} method}, establishing an analytical framework directly in the image domain (Section~\ref{Experimental Design}).
Specifically, we demonstrate that the complex spatial filtering and instrumental effects of an interferometer can be analytically encapsulated into a single explicit equation:
$I_{\mathrm{pred}} = \sum_{l=1}^{n} [\mathrm{CDD}_l(I_{\mathrm{in}}) \times R(l)] + N_{\mathrm{rms}}$
where $\mathrm{CDD}_l(I_{\mathrm{in}})$ represents the multiscale components decomposed from the true sky brightness, and $N_{\mathrm{rms}}$ denotes the synthesized instrumental thermal noise.
Crucially, $R(l)$ is a one-dimensional error function (\texttt{erf}) transfer weight that quantifies the flux recovery fraction at scale $l$ (Section~\ref{results:erf_fits}); its spatial filtering cutoff is physically governed by the array's theoretical MRS $\theta_{\mathrm{MRS, theoretical}} \approx 0.6 \lambda / L_{\mathrm{min}}$), directly binding the mathematical prediction to the minimum baseline length ($L_{\mathrm{min}}$).
By applying this comprehensive analytical formula, the CDD-\texttt{erf} framework allows for the instantaneous generation of realistic mock observations (Section~\ref{results:cdd_erf_prediction}).
It effectively predicts the missing flux as a continuous function of spatial scale, completely bypassing the need for computationally expensive Fourier-domain simulations.
\section{Experimental Design}\label{Experimental Design}
\begin{figure}[H]
\centering
\includegraphics[width=1\textwidth]{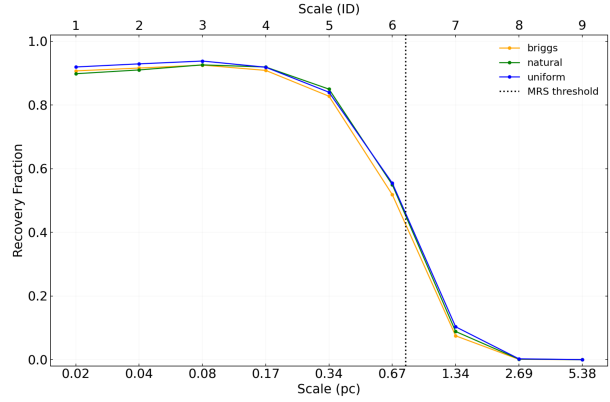}
\caption{\textbf{Workflow for predicting interferometric images from multiscale components.}
Step~1: Forward modeling with \textsc{casa} shows the loss of large-scale emission due to spatial filtering.
Step~2: The simulated data is decomposed into scale-space using the CDD method.
Step~3: The flux recovery fraction as a function of spatial scale is described by a one-dimensional \texttt{erf}.
Step~4: The \texttt{erf} transfer function is applied to the scale-space
components to produce the predicted interferometric image, bypassing full visibility simulations.
}
\label{fig1:workflow}
\end{figure}

As outlined in the Introduction, the core of our CDD-\texttt{erf} method predictive framework is an explicit analytical mapping from the true sky brightness ($I_{\mathrm{in}}$) to the final predicted spatially filtered interferometric observation ($I_{\mathrm{pred}}$).
Formally, this is governed by the equation:
\begin{equation} \label{eq:master_equation}
    I_{\mathrm{pred}} = \sum_{l=1}^{n} \Big[ \mathrm{CDD}_l(I_{\mathrm{in}}) \times R(l) \Big] + N_{\mathrm{rms}},
\end{equation}
\begin{equation} \label{eq:erf_equation}
    R(l) = \frac{B}{2} \left[ 1 - \mathrm{erf}\left( \frac{l - c_{\mathrm{recover}}}{w} \right) \right],
\end{equation}
where $\mathrm{CDD}_l$ denotes the $l$-th scale component extracted by the CDD method, and $R(l)$ is the scale-dependent transfer function modeled by a one-dimensional \texttt{erf},
with $l$ being the scale index, $B$ the maximum recovery fraction, $w$ the transition width, and $c_{\mathrm{recover}}$ the spatial filtering cutoff.
This mathematical cutoff $c_{\mathrm{recover}}$ is explicitly tied to the physical properties of the interferometric array via $c_{\mathrm{recover}} = \log_2(\theta_{\mathrm{MRS}} / \theta_{\mathrm{pixel}})$, where $\theta_{\mathrm{pixel}}$ is the pixel scale of the image.
The $\theta_{\mathrm{MRS}}$ is fundamentally governed by the minimum baseline length ($L_{\mathrm{min}}$) of the array at observing wavelength $\lambda$, given by the theoretical relation $\theta_{\mathrm{MRS, theoretical}} \approx 0.6 \lambda / L_{\mathrm{min}}$ \citep{cortes_2025_14933753}.
This direct linkage grounds our analytical master equation to actual instrumental hardware limits.
The term $N_{\mathrm{rms}}$ represents the instrument's thermal noise, generated as a two-dimensional Gaussian white noise field, $\mathcal{N}(0, \sigma_{\mathrm{rms}}^2)$.
The theoretical root-mean-square (RMS) noise level, $\sigma_{\mathrm{rms}}$, is determined by the standard point-source sensitivity equation for interferometric arrays \citep{cortes_2025_14933753}.
In our framework, $N_{\mathrm{rms}}$ serves as a user-defined input parameter to match specific observational requirements.
To translate this theoretical formulation into a practical predictive tool, we must empirically determine the exact behavior of the transfer function $R(l)$ and validate the framework's overall fidelity.
To achieve this, we design a systematic four-step workflow based on standard mock observations, as illustrated in Figure~\ref{fig1:workflow}:
\begin{itemize}
    \item \textbf{Step 1: Mock observation.}
We simulate the target field using the \textsc{casa} simulator.
The procedure consists of preparing a sky model image, simulating visibilities, imaging, and combining the resulting data products.
We produce interferometric ($12\,\mathrm{m} + 7\,\mathrm{m}$) and combined ($12\,\mathrm{m} + 7\,\mathrm{m} + \mathrm{TP}$) intensity maps in units of $\mathrm{Jy\,beam^{-1}}$.
The simulation procedure, parameter settings, and spatial scale matching are described in Section~\ref{sec:simulation}.
    \item \textbf{Step 2: Multi-scale decomposition.}
We apply the CDD method to the mass surface density maps derived from the simulated ALMA $3\,\mathrm{mm}$ continuum intensity maps.
The input image ($I_{\mathrm{in}}$) is decomposed into $n$ scale components using the CDD method:
\begin{equation}
    I_l = \mathrm{CDD}_l(I_{\mathrm{in}}), \quad l \in [1, n].
\end{equation}
This produces a set of multiscale component maps ($I_1, I_2, \ldots, I_n$), where each component $I_n$ isolates structures with angular sizes in the range $[2^n, 2^{n+1}]$ pixels, together with a residual map containing all background emission on scales larger than the maximum decomposition scale.
From these component maps, we quantify key physical properties --- including structural morphology, flux distributions, flux recovery fractions, and mass fractions --- as a function of spatial scale.
    \item \textbf{Step 3: Scale-dependent transfer function.}
From the CDD analysis in Step 2, we extract the flux recovery fraction as a continuous function of spatial scale.
As introduced in Equation~(\ref{eq:erf_equation}), we establish that this scale-dependent recovery curve can be robustly modeled by the one-dimensional \texttt{erf} transfer function, from which we determine the best-fit parameters ($B$, $w$, and $c_{\mathrm{recover}}$) (Section~\ref{results:erf_fits}).
    \item\textbf{Step4: Scale-informed image prediction.}
Having established that the spatial filtering of an interferometer can be mathematically described by the \texttt{erf} function in the CDD scale-space, we can predict the observational outcome directly in the image domain.
As encapsulated by the equation (Equation~\ref{eq:master_equation}), the predicted interferometric components are obtained by multiplying each decomposed scale by its \texttt{erf} weight, linearly recombining them, and injecting the theoretical thermal noise $N_{\mathrm{rms}}$.
This provides a rapid, explicit-form analytical mapping from the true sky to the final interferometric observation (Section~\ref{results:cdd_erf_prediction}).
\end{itemize}
\vspace{-20pt}
\section{Data}
\subsection{Data Simulation} \label{sec:simulation}
The Perseus molecular cloud has been extensively studied through distance and kinematic surveys \citep{2010A&A...512A..67L, 2018ApJ...865...73O}, molecular line emission \citep{2006AJ....131.2921R, 2024ApJ...977..135C, 2024MNRAS.528..577M}, dust continuum observations \citep{2006ApJ...638..293E, 2016A&A...587A.106Z, 2022RAA....22e5012Z}, high-resolution protostellar surveys \citep{2016ApJ...818...73T}, and three-dimensional magnetic field morphology \citep{2022A&A...660A..97T}.
Its three-dimensional physical structure \citep{2021ApJ...919L...5B, 2021ApJ...919...35Z} and three-dimensional density exponents \citep{2022MNRAS.514L..16L} have also been characterized.
The Perseus dataset has furthermore served as a testbed for new analytical techniques,
including the G-virial gravity-based structure analysis \citep{2015A&A...578A..97L, 2016arXiv160305720L} and the CDD multiscale decomposition method \citep{2022ApJS..259...59L}.
This extensive observational and theoretical foundation provides robust constraints on key physical parameters, including distance, optical depth, dust temperature, extinction, column density, star-formation efficiency, and catalogs of young stellar objects (YSOs).
For these reasons, we selected the Perseus molecular cloud as the target region.
We divided the cloud into six sub-regions --- IC~348, L1455, B1, NGC~1333, L1448, and the Perseus Cirrus --- and processed each independently.
We used \textit{Herschel}/SPIRE $500\,\mu\mathrm{m}$ observations (see Figure~\ref{fig:appendix_map}) \footnote{Data retrieved from the NASA/IPAC Infrared Science Archive (IRSA): \url{https://irsa.ipac.caltech.edu/}} as the input sky model for the \textsc{casa} simulator to simulate ALMA $3\,\mathrm{mm}$ continuum emission.
The simulation consisted of two main steps: visibility generation, and imaging with data combination.
For the visibility generation, we produced the interferometric $uv$ data using the \texttt{simobserve} task in \textsc{casa}.
The simulations model ALMA Band~3 continuum emission with a central frequency of $87.515\,\mathrm{GHz}$ and a bandwidth of $937.5\,\mathrm{MHz}$.
Because each sub-region spans approximately $1\degr \times 1\degr$ on the sky, simulating the full field at its native angular scale is computationally prohibitive.
We therefore rescaled the angular pixel size of the input model by modifying the \texttt{CDELT} header keyword, effectively simulating the source as if observed at a greater distance.
This rescaling enables efficient mosaic coverage.
We applied three scaling factors --- $0.10$, $0.07$, and $0.04$ --- yielding effective pixel sizes of approximately $0\farcs90$, $0\farcs63$, and $0\farcs36$, respectively, corresponding to the following array configurations:
\begin{itemize}
    \item \textbf{C-1 configuration:} scaling factor $0.10$ (hereafter \texttt{far0.1}), angular resolution $\sim\!3\farcs19$, $\mathrm{MRS} \sim\!28\farcs5$.
    \item \textbf{C-3 configuration:} scaling factor $0.07$ (hereafter \texttt{far0.07}), angular resolution $\sim\!2\farcs15$, $\mathrm{MRS} \sim\!16\farcs2$.
    \item \textbf{C-4 configuration:} scaling factor $0.04$ (hereafter \texttt{far0.04}), angular resolution $\sim\!1\farcs30$, $\mathrm{MRS} \sim\!11\farcs2$.
\end{itemize}
We adopted the ALMA Cycle~12 antenna configuration files \texttt{alma.cycle12.1.cfg}, \texttt{alma.cycle12.3.cfg}, and \texttt{alma.cycle12.4.cfg} for the $12\,\mathrm{m}$ array, \texttt{aca.cycle12.cfg} for the $7\,\mathrm{m}$ Atacama Compact Array (ACA; angular resolution $\sim\!12\farcs7$, $\mathrm{MRS} \sim\!66\farcs7$), and \texttt{aca.tp.cfg} for total power (TP) observations.
The total integration times were $2\,\mathrm{hr}$ for the $12\,\mathrm{m}$ array, $4\,\mathrm{hr}$ for the $7\,\mathrm{m}$ array, and $8\,\mathrm{hr}$ for the TP array, with an integration time of $10\,\mathrm{s}$ per pointing.
Atmospheric conditions were modeled adopting a precipitable water vapor column of $0.5\,\mathrm{mm}$.
For the imaging and data combination step, we imaged the simulated visibilities using the \texttt{tclean} task.
To assess the sensitivity of our results to different density regimes, we tested three visibility weighting schemes: natural, Briggs (robust parameter $= 0.5$), and uniform.
The $12\,\mathrm{m}$ and $7\,\mathrm{m}$ interferometric data were combined in the visibility domain using the \texttt{concat} task.
The TP data were subsequently combined with the interferometric maps in the image plane using the \texttt{feather} task.
This procedure yielded the final interferometric ($12\,\mathrm{m} + 7\,\mathrm{m}$; hereafter ``inte'') and combined ($12\,\mathrm{m} + 7\,\mathrm{m} + \mathrm{TP}$; hereafter ``comb'') intensity maps in units of $\mathrm{Jy\,beam^{-1}}$ for each array configuration and weighting scheme.
\vspace{-5pt}
\subsection{Grid Alignment and Flux Calibration}\label{data:Harmonizing}
Before analysis, we applied a renormalization step to the output images to bring all simulation sets onto a common spatial grid.
The simulations were carried out at rescaled angular pixel sizes of $0\farcs90$, $0\farcs63$, and $0\farcs36$ for the \texttt{far0.1}, \texttt{far0.07}, and \texttt{far0.04} configurations, respectively.
We restored the pixel scale in all output FITS images to the native resolution of the input sky model ($9\arcsec\,\mathrm{pixel}^{-1}$; corresponding to a physical scale of $0.0105\,\mathrm{pc\,pixel}^{-1}$ assuming a distance of $240\,\mathrm{pc}$ to the Perseus molecular cloud; \citealt{2010A&A...512A..67L}).
By mapping all simulation sets back onto this common $9\arcsec$ grid, we ensure that a given scale index corresponds to the same physical structure across all configurations.
Any differences in the flux recovery fraction among configurations can therefore be attributed solely to differences in the interferometric spatial filtering response, rather than to inconsistencies in the analysis grid.
Although the input sky model is based on $500\,\mu\mathrm{m}$ intensity, which differs in absolute flux scale from the simulated $3\,\mathrm{mm}$ emission, the relative spatial structure is representative of optically thin dust emission at millimeter wavelengths.
The CDD-based analysis of the structural distribution across spatial scales therefore remains physically valid.
Furthermore, the mass recovery fraction $M_{12\,\mathrm{m}+7\,\mathrm{m}} /
M_{12\,\mathrm{m}+7\,\mathrm{m}+\mathrm{TP}}$ derived from the CDD decomposition is inherently a ratio quantity, and its quantitative validity is unaffected by the absolute flux normalization.
Because this study focuses on the recovery fraction and structural analysis, the absolute flux scaling cancels and does not influence our conclusions regarding the interferometric response.
Assuming optically thin dust emission, the observed intensity is related to the gas mass surface density $\Sigma_{\mathrm{gas}}$ by $I_\nu = B_\nu(T_d) (\kappa_\nu / R_{\mathrm{gd}}) \Sigma_{\mathrm{gas}}$ where $B_\nu(T_d)$ is the Planck function evaluated at dust temperature $T_d$, $\kappa_\nu$ is the dust opacity, and $R_{\mathrm{gd}} = 100$ is the assumed gas-to-dust mass ratio.
To place the simulated maps on a physically meaningful surface density scale, we apply a theoretically derived flux conversion factor $[C = I_{500\,\mu\mathrm{m}}/I_{3\,\mathrm{mm}} = \kappa_{500\,\mu\mathrm{m}} B_{500\,\mu\mathrm{m}}(T_d) / (\kappa_{3\,\mathrm{mm}} B_{3\,\mathrm{mm}}(T_d)) \approx 726]$, computed adopting dust properties characteristic of Perseus ($T_d = 20\,\mathrm{K}$, emissivity spectral index $\beta = 1.8$; e.g., \citealt{2016A&A...587A.106Z}).
This factor corrects for the spectral energy distribution difference between the two wavelengths and anchors the simulated intensities to physical mass surface densities.
\section{Model Calibration}\label{sec:results}
In this section, we calibrate and validate our analytical framework against the \textsc{casa} simulations.
As established, the final calibrated predictive workflow is governed by the equation:
\begin{equation}
    I_{\mathrm{pred, calibrated}} = k \times \left( \sum_{l=1}^{n} \Big[ \mathrm{CDD}_l(I_{\mathrm{in}}) \times R(l) \Big] \right) + N_{\mathrm{rms}}.
\end{equation}
To systematically build, verify, and apply this equation, the remainder of this section is organized into four logical steps.
First, we characterize the missing flux in scale-space and determine the transfer function $R(l)$ driven by the array's physical baseline limits (Section~\ref{results:erf_fits}).
Next, we apply this analytical model to predict interferometric images, evaluating their global mass recovery and morphological fidelity (Section~\ref{results:cdd_erf_prediction}).
Then, to correct for absolute flux offsets introduced by non-linear algorithmic biases, we derive and apply the pixel-by-pixel alignment factor $k$ (Section~\ref{results:alignment_factor}).
Finally, having established a robust and fully calibrated framework, we utilize it to investigate the underlying astrophysical drivers, demonstrating that interferometric flux loss is fundamentally dictated by spatial scale rather than gas surface density (Section~\ref{results:physical_driver}).
\vspace{-3pt}
\subsection{Characterizing Missing Flux in Scale-Space}\label{results:erf_fits}
\begin{figure}[ht!]
\centering
\includegraphics[width=1\textwidth]{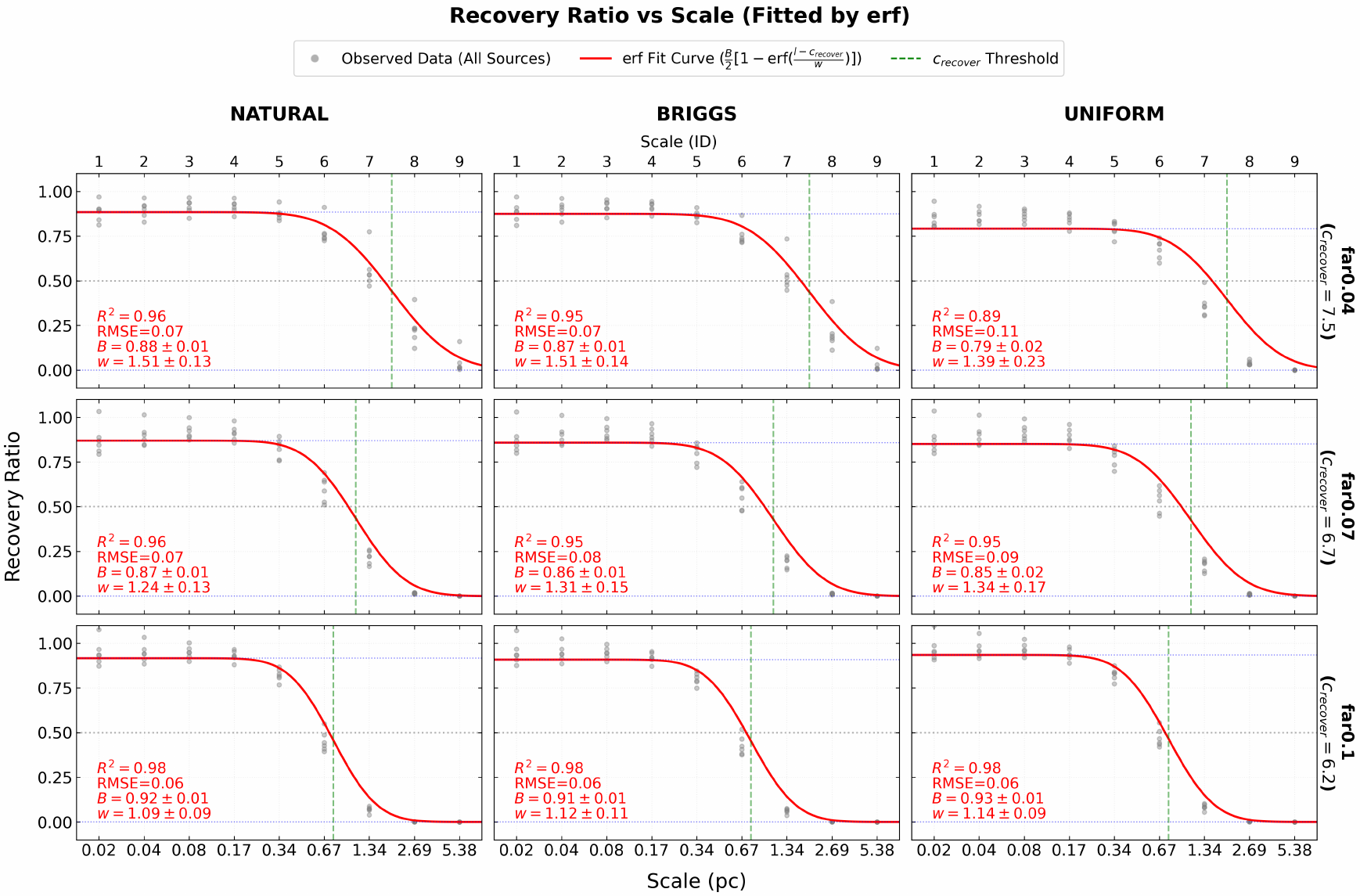}
\caption{\textbf{Error function fits to the flux recovery fraction as a function of spatial scale for all sub-regions.}
Gray points show the measured recovery fractions.
The red curve shows the \texttt{erf} fit.
The vertical green dashed line marks the theoretical scale-index cutoff ($c_{\mathrm{recover}}$, indicated in the right-hand legend) corresponding to the MRS of the ACA $7\,\mathrm{m}$ array ($66\farcs7$) for each configuration, which is fixed as an input rather than fitted.
The coefficient of determination ($R^2$), the root-mean-square error (RMSE), and the best-fit parameters ($B$ and $w$) along with their standard errors are listed in the lower-left corner of each panel.
Rows correspond to different array configurations (\texttt{far0.04}, \texttt{far0.07}, and \texttt{far0.1}).
Columns correspond to different imaging weighting schemes (natural, Briggs, and uniform).
}
\label{fig2:Erf_Fit}
\end{figure}

Fundamentally, scale space is a method of presenting an image as a set of sub-images, distinguished by their respective spatial scales.
The modern mathematical foundation of scale-space theory was heavily shaped by Tony Lindeberg \citep[e.g.,][]{1994JApSt..21..225L,1994sstc.book.....L}, who systematically formalized how image features naturally appear, persist, and disappear across different levels of resolution without introducing spurious artifacts.
Unlike classic scale-space theory based on linear Gaussian blurring, the CDD framework uses anisotropic, non-linear diffusion driven by local image contrast and geometric features \citep{2022ApJS..259...59L}.
This approach selectively smooths noise while actively preserving prominent object boundaries, effectively decomposing an input image into a set of distinct multiscale sub-images.
A visual comparison of these decomposed components (Figure~\ref{fig1:workflow}, Step~2) reveals that the interferometric spatial filtering effect is strictly scale-dependent.
While compact structures ($I_1$ through $I_6$) are morphologically consistent between the interferometric and combined maps, a systematic divergence emerges at larger scales ($I_7, I_8$, and the residual $I_9$).
For these extended envelopes, the interferometric emission progressively decreases to zero (producing characteristic negative bowls), whereas the combined data successfully recovers the diffuse background.
Quantitatively, this scale-dependent flux loss of interferometric spatial filtering is intrinsically governed by the array's minimum baseline, $L_{\mathrm{min}}$.
The absolute physical upper boundary for structural recovery is dictated by the theoretical MRS, $\theta_{\mathrm{MRS, theoretical}} \approx 0.6\,\lambda/L_{\mathrm{min}}$, arising from the unsampled origin of the $uv$-plane.
To ensure robust imaging quality, ALMA adopts a more conservative operational limit using the 5th-percentile baseline: $\theta_{\mathrm{MRS,\, operational}} \approx 0.983\,\lambda / L_{5}$ \citep{cortes_2025_14933753}.
Projected into our CDD scale-space, this physical limit defines an effective scale-index cutoff, $c_{\mathrm{recover}} = \log_2(\theta_{\mathrm{MRS}} / \theta_{\mathrm{pixel}})$.
Consequently, the cutoff indices for the \texttt{far0.1}, \texttt{far0.07}, and \texttt{far0.04} configurations are approximately $6.2$, $6.7$, and $7.5$, respectively.
Beyond this absolute MRS threshold, the interferometer cannot recover emission, regardless of the imaging weighting scheme (which only acts as a secondary regulator for scales within the MRS).
While standard interferometric imaging (e.g., \textsc{clean}) constitutes a complex non-linear pipeline, this spatial filtering takes a tractable mathematical form in the CDD scale-space.
By computing the mass recovery fraction ($M_{\mathrm{inte}} / M_{\mathrm{comb}}$) across scales (Figure~\ref{fig2:Erf_Fit}), we find that the attenuation consistently exhibits a sigmoid-like profile.
This behavior---a flat plateau ($\gtrsim 0.85$ recovery) at small scales followed by a steep decline beyond the MRS threshold---can be explicitly captured by a one-dimensional \texttt{erf} model (Equation~\ref{eq:erf_equation}), mathematically representing the $1/e$ attenuation transition across the baseline limits.
As demonstrated in Figure~\ref{fig2:Erf_Fit}, the \texttt{erf} model accurately fits the empirical recovery fractions.
To evaluate the goodness of fit, we calculated the coefficient of determination ($R^2$) and the root-mean-square error (RMSE) for all regressions.
Across the different configurations and weighting schemes, the fits yield $R^2$ values ranging from $0.89$ to $0.98$ and RMSE values between $0.06$ and $0.11$.
These metrics indicate that the continuous one-dimensional \texttt{erf} function provides a reliable empirical framework for describing the spatial filtering response.

It is crucial to emphasize that the cutoff index $c_{\mathrm{recover}}$ is not a free parameter in the fitting.
Instead, it is strictly fixed to the theoretical value derived from the array's minimum baseline ($\theta_{\mathrm{MRS}}$).
Consequently, the \texttt{erf} fitting only determines two free parameters: the asymptotic upper limit $B$ and the transition width $w$.
These fitted parameters correspond to distinct physical properties of the interferometric observations.
First, the asymptotic upper limit, $B$, which represents the maximum recoverable fraction for compact structures, yields values of $0.85$--$0.95$, with standard errors (SE) $\lesssim 0.02$.
This indicates that compact structures exhibit a systematic flux deficit in standard interferometric imaging, although this loss can be constrained to $<15\%$ depending on the visibility weighting.
Under uniform weighting combined with the extended \texttt{far0.04} configuration, $B$ decreases to $\sim 0.79$.
This reflects that the suppression of short baselines associated with resolution enhancement causes the interferometer to partially resolve out the extended envelopes of the compact cores.
Furthermore, individual data points in Figure~\ref{fig2:Erf_Fit} exceed a recovery fraction of $1.0$, which occurs because the non-linear \textsc{clean} algorithm sharpens unresolved peaks to compensate for negative bowls.
The convergence of $B$ to values $<1$ and the SE of $\lesssim 0.02$ confirm that the \texttt{erf} model captures the recovery trend without being biased by these outliers.

Second, the transition width $w$ depends on the array configuration relative to the angular scale of the source.
As the source distance increases ($w_{\mathrm{far,0.04}} > w_{\mathrm{far,0.07}} > w_{\mathrm{far,0.1}}$), the extended envelopes subtend smaller angular scales, shifting their spatial frequency content toward the short-baseline regime of the ACA $7\,\mathrm{m}$ array, where the $uv$-plane is more densely sampled.
Consequently, a larger fraction of the source flux is recovered, resulting in a broader transition width $w$.
The SE of the fitted parameters are $\Delta w \sim 0.09$--0.23, indicating that this trend remains present when parameter uncertainties are included.

Conversely, for extended configurations, large-scale structures fall directly into the zero-spacing gap where the $uv$-plane is sparsely sampled, producing a step-like cutoff and a smaller $w$.
These results demonstrate that interferometric spatial filtering can be analytically decoupled and expressed as an explicit mathematical transfer function.
The fact that the fixed $c_{\mathrm{recover}}$ (derived from the array's physical baseline) generates such robust \texttt{erf} fits raises a fundamental question: why does a non-linear CDD image-domain scale map effectively to a Fourier-domain instrumental cutoff?
Addressing this question requires clarifying the relationship between the CDD scale and the Fourier-domain spatial frequency.
These two concepts are mathematically distinct. In standard Fourier analysis, spatial frequency is a global metric defined as the inverse of the angular scale of the Fourier basis functions. In contrast, the CDD framework provides a localized definition of spatial scale driven by constrained diffusion. Specifically, a scale in CDD relates to the effective width (dispersion) of the Gaussian kernel applied at each step of the diffusion process \citep{2022ApJS..259...59L}.
Unlike linear Gaussian blurring, this non-linear diffusion allows the CDD framework to adapt to hierarchical structures in astronomical images while preserving non-negative flux and avoiding the ringing artifacts of global Fourier bandpass filters.

Mapping the $uv$-domain instrumental limit ($\theta_{\mathrm{MRS}}$) to a CDD scale cutoff ($c_{\mathrm{recover}}$) suggests an underlying physical correspondence.
From a signal processing perspective, a localized physical structure with a characteristic angular size $\theta$ concentrates its Fourier power at a spatial frequency $u \sim 1/\theta$.
Furthermore, because molecular clouds exhibit continuous, hierarchical density distributions, this local-to-global scale relationship applies across the dynamic range.
This statistical matching, rather than mathematical equivalence, underlies the observed empirical correspondence.
Thus, while the CDD scale is not the mathematical inverse of the spatial frequency, there is a monotonic correlation between them.
This structural correspondence justifies embedding the physical $\theta_{\mathrm{MRS}}$ into the scale-space \texttt{erf} model as an observational proxy.

\vspace{-3pt}
\subsection{High-Fidelity Reconstruction and Prediction}\label{results:cdd_erf_prediction}
\begin{figure}[ht!]
\centering
\includegraphics[width=1\textwidth]{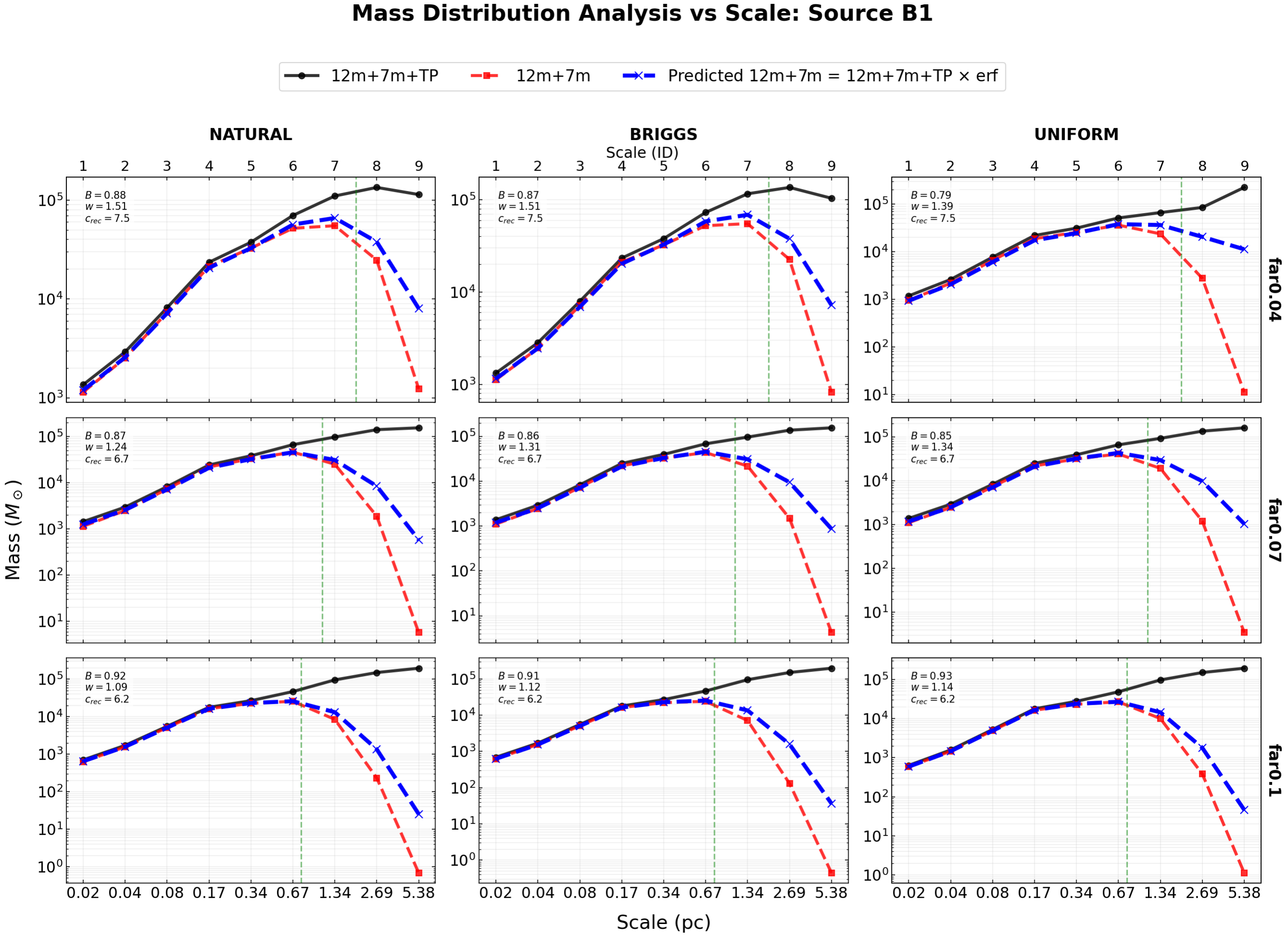}
\caption{\textbf{Comparison of the observed and predicted mass as a function of spatial scale.}
The black solid, red dashed, and blue dashed lines show the combined data, the simulated interferometric data, and the predicted interferometric data, respectively.
The predicted values are obtained by applying the \texttt{erf} fit
(Figure~\ref{fig2:Erf_Fit}) to the combined data.
The vertical axis shows the mass contained in each scale component. The top and bottom horizontal axes show the scale index and the corresponding physical scale (pc), respectively.
The best-fit \texttt{erf} parameters are listed in the upper-left corner of each panel.
The vertical green dashed line, top and bottom horizontal axes, and all other aspects of the panel layout follow Figure~\ref{fig2:Erf_Fit}.
}
\label{fig3:mass_vs_scale}
\end{figure}

\begin{figure}[ht!]
\centering
\includegraphics[width=0.95\textwidth]{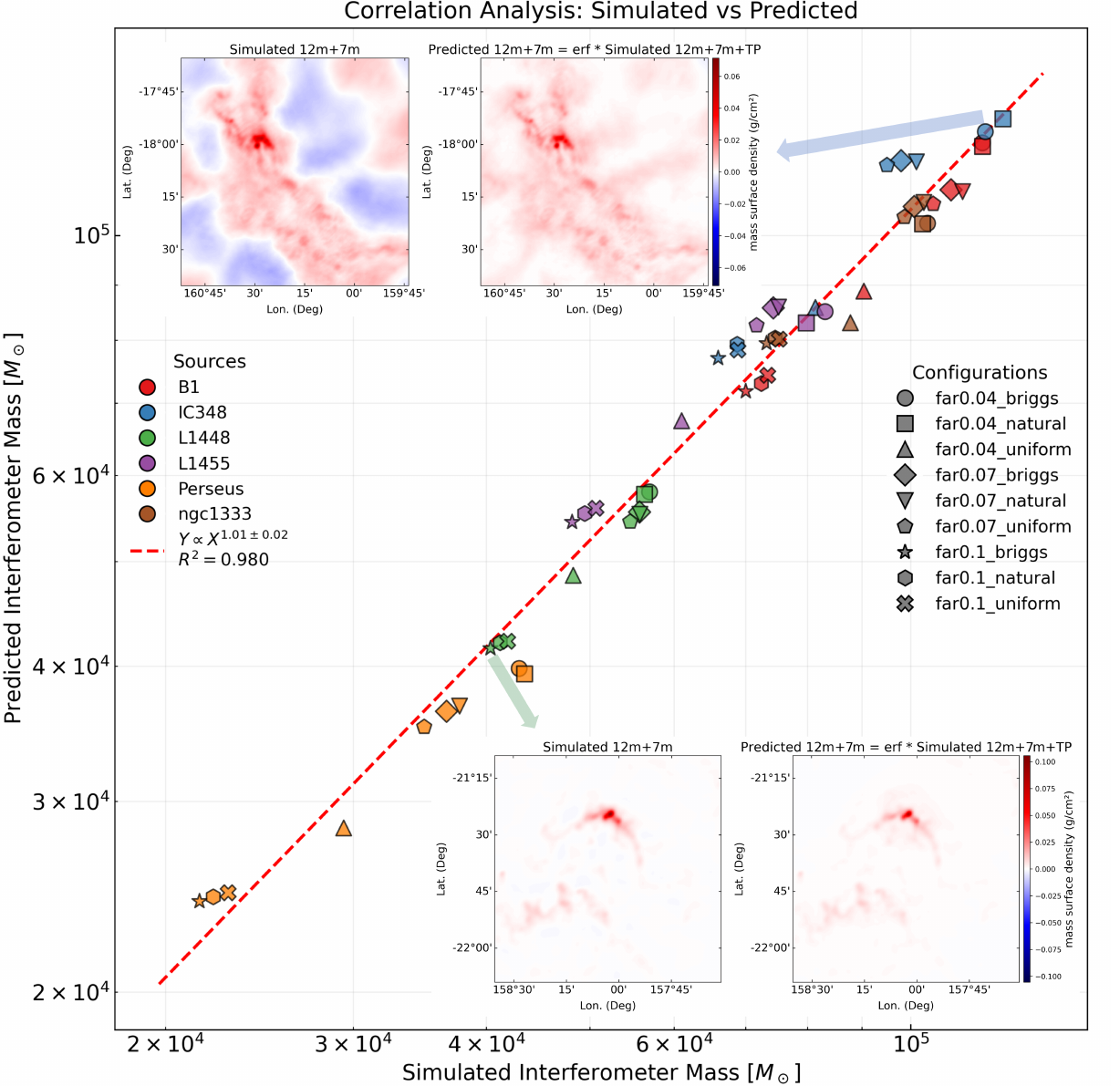}
\caption{\textbf{Predicted versus simulated interferometric mass.}
The horizontal and vertical axes show the \textsc{casa}-simulated and CDD--\texttt{erf}-predicted interferometric masses, respectively, integrated within the $c_{\mathrm{recover}}$ cutoff.
Different colors indicate different target sources, and different symbols denote different array configurations.
The red dashed line shows the unconstrained ordinary least-squares fit performed in logarithmic space,
$\log \rm Y = a + b \log \rm X$.
The best-fit relation is consistent with a nearly linear scaling, $\rm Y \propto \rm X^{1.01 \pm 0.02}$, with $R^2 = 0.980$.
The insets display the \textsc{casa}-simulated and CDD--\texttt{erf}-predicted $12\,\mathrm{m} + 7\,\mathrm{m}$ mass surface density maps for two of the sources.
}
\label{fig4:erf_predict}
\end{figure}

Having established that the spatial filtering of an interferometer can be described mathematically by the \texttt{erf} function in the CDD scale-space, we predict the interferometric flux distribution by applying Equation~\ref{eq:erf_equation}.
Figure~\ref{fig3:mass_vs_scale} compares the absolute mass distributions as a function of spatial scale.
The predicted interferometric mass (blue dashed line) closely follows the \textsc{casa}-simulated interferometric mass (red dashed line).
For compact components ($I_1$ through $I_5$), the predicted and simulated curves are perfectly consistent with the total combined mass.
At scales exceeding the MRS cutoff (green vertical line), both curves diverge from the total mass and exhibit identical attenuation profiles, eventually falling to a negligible level.
The fidelity of this prediction confirms that our analytical framework accurately reproduces the scale-dependent filtering behavior.
Furthermore, the \texttt{erf} correctly captures the shift of the MRS threshold ($c_{\mathrm{recover}}$) toward smaller spatial scales as the array configuration becomes more extended, successfully predicting the corresponding decrease in recovered mass at large scale indices.
To assess the global predictive power of the model, Figure~\ref{fig4:erf_predict} presents a correlation analysis between the total predicted mass and the actual simulated interferometric mass across all 54 parameter combinations (6 sub-regions $\times$ 3 distances $\times$ 3 weighting schemes).
For each sub-region, the total interferometric mass is computed by integrating only over spatial scales $\leq \mathrm{MRS}$, defining a high-fidelity recoverable mass that represents the physical structures the interferometer can reliably detect.
The data points are well described by an unconstrained ordinary least-squares fit performed in logarithmic space,
\[
\log \rm Y = a + b \log \rm X,
\]
where $\rm X$ and $\rm Y$ denote the simulated and predicted interferometric masses, respectively, and both $a$ and $b$ are free parameters.
The best-fit relation is consistent with a nearly linear scaling, $\rm Y \propto \rm X^{1.01 \pm 0.02}$, with $R^2 = 0.980$.
This near-unity slope indicates that, within the recoverable-scale mass budget defined by the $c_{\mathrm{recover}}$ cutoff, the CDD--\texttt{erf} method reproduces the \textsc{casa}-simulated interferometric masses with little systematic scale-dependent bias.

\begin{figure}[ht!]
\centering
\includegraphics[width=1\textwidth]{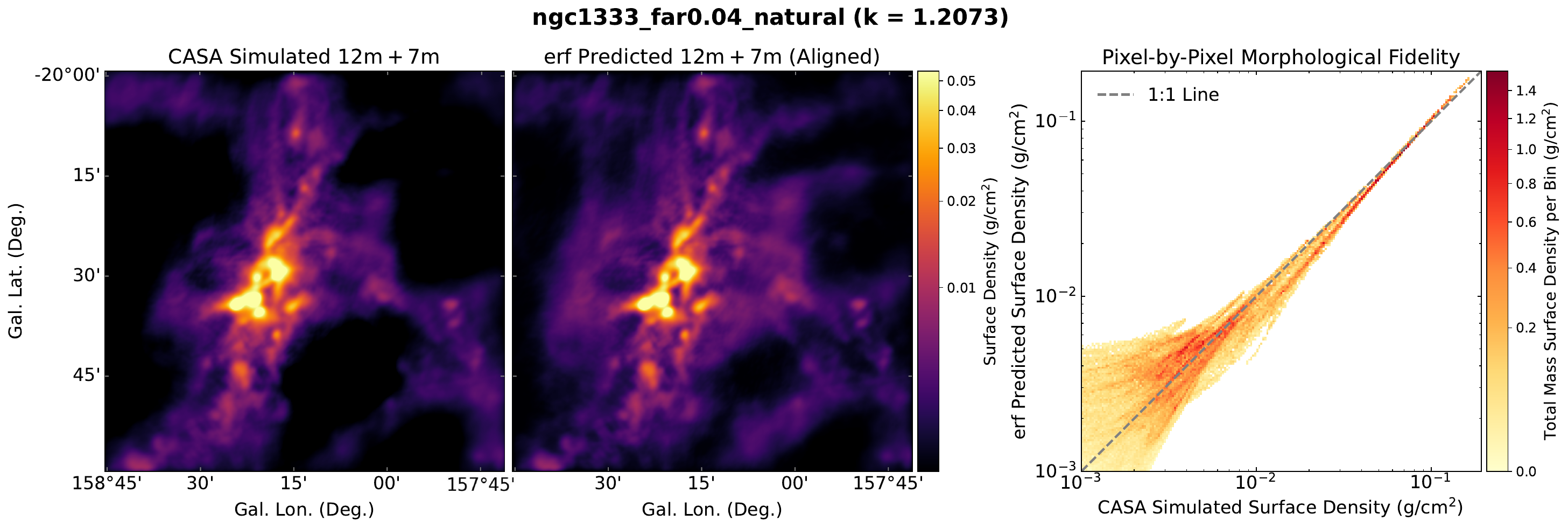}
\caption{\textbf{Morphological reconstruction of the NGC\,1333 sub-region under the \texttt{far0.04} configuration with natural weighting.}
The left and center panels show the \textsc{casa}-simulated $12\,\mathrm{m}+7\,\mathrm{m}$ mass surface density map
and the CDD--\texttt{erf}-predicted map after applying the mass-weighted mean alignment factor ($k$), respectively.
The right panel shows the corresponding pixel-by-pixel two-dimensional histogram.
The horizontal and vertical axes represent the value of \textsc{casa} simulations and the aligned predictions, respectively, with both axes restricted to $\Sigma > 10^{-3}\,\mathrm{g\,cm^{-2}}$.
The color scale indicates the total mass summed within each bin (note that this is heavily bin-size dependent and does not represent an independent physical quantity).
The gray dashed line indicates a slope of 1.
}
\label{fig5:example_ngc1333}
\end{figure}

This agreement confirms that the CDD--\texttt{erf} method prediction is highly robust across diverse molecular cloud morphologies.
Beyond integrated mass, spatial morphology is an essential diagnostic in astrophysical analysis.
Figure~\ref{fig5:example_ngc1333} shows a representative case study of the NGC\,1333 sub-region (analogous figures for the remaining sub-regions are provided in Appendix \ref{fig:appendix_other_sources}).
The predicted map (center panel) is morphologically highly consistent with the \textsc{casa}-simulated map (left panel): the dense protostellar cores and filamentary structures are beautifully reproduced, while the extended diffuse emission is systematically resolved out.
The pixel-by-pixel histogram (right panel) visually confirms this excellent structural correlation across the dynamic range of the map.
Note that to achieve the tight 1:1 absolute flux alignment shown in the center and right panels, the raw analytical prediction was globally scaled by a configuration-specific alignment factor ($k$).
The rigorous derivation and physical necessity of this absolute flux calibration are detailed in the subsequent section (Section~\ref{results:alignment_factor}).
\subsection{Pixel-by-Pixel Alignment Factor}\label{results:alignment_factor}
\begin{figure}[ht!]
\centering
\includegraphics[width=1\textwidth]{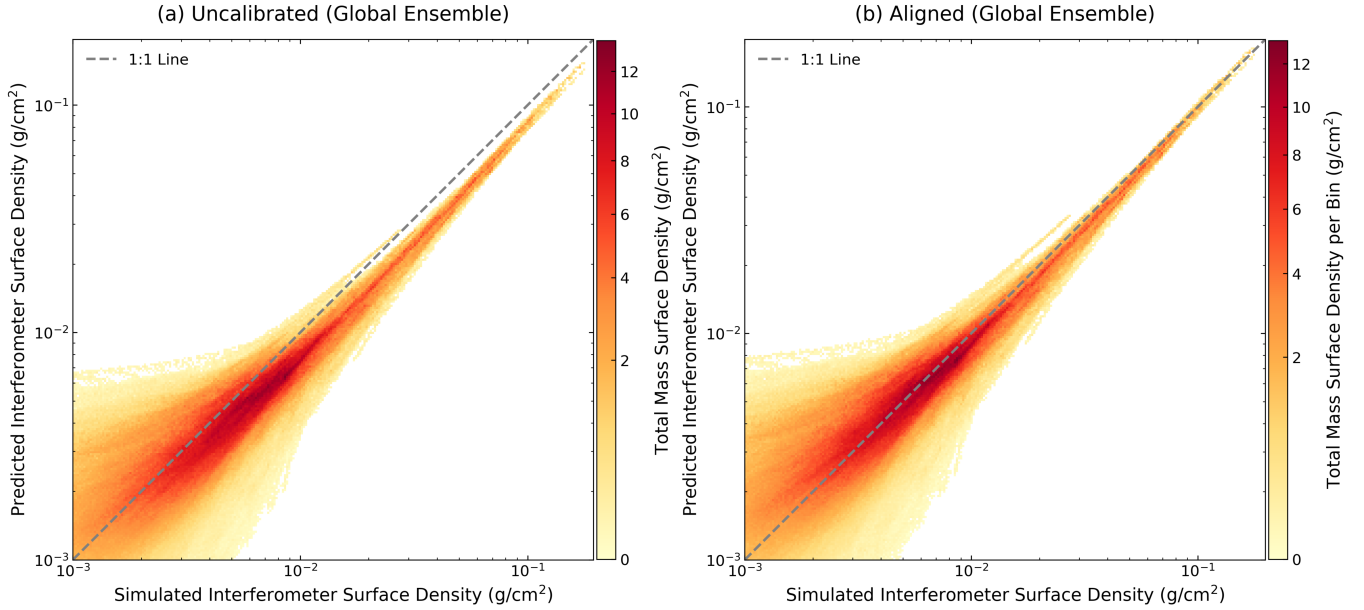}
\caption{\textbf{Global pixel-by-pixel fidelity of the CDD--\texttt{erf} method before (a) and after (b) mass-weighted mean alignment.}
These two-dimensional histograms demonstrate the predictive fidelity of our method across all 54 mock observational configurations.
Axis conventions and pixel selection criteria are identical to those in the right panel of Figure~\ref{fig5:example_ngc1333}
}
\label{fig6:2d_hist}
\end{figure}

While the multiscale spatial morphology is accurately predicted by the analytical transfer function (as demonstrated in Figure~\ref{fig5:example_ngc1333}), robust astrophysical analysis requires strict absolute flux fidelity.
It is well established that non-linear deconvolution algorithms, such as the \textsc{casa} \texttt{tclean} task, can artificially sharpen unresolved peak intensities and introduce absolute flux calibration offsets relative to pure analytical linear models \citep{2024MNRAS.527..942R,2025ASPC..541..155W}.
To correct for this algorithmic discrepancy and achieve rigorous absolute flux matching, we apply a configuration-specific mass-weighted mean alignment factor ($k$), e.g., on a pixel-by-pixel basis, $I_{\mathrm{pred, calibrated}} = k I_{\mathrm{pred}}$.
And to be more accurate, we define the mass-weighted mean surface densities for the \textsc{casa}-simulated maps ($\bar{\Sigma}_{\mathrm{inte}}$) and the uncalibrated CDD--\texttt{erf}-predicted maps ($\bar{\Sigma}_{\mathrm{pred}}$) as:
\begin{equation}
    \bar{\Sigma}_{\mathrm{inte}} = \frac{\sum_{i} \Sigma_{\mathrm{comb},\,i}\,
    \Sigma_{\mathrm{inte},\,i}}{\sum_{i} \Sigma_{\mathrm{comb},\,i}}, \qquad
    \bar{\Sigma}_{\mathrm{pred}} = \frac{\sum_{i} \Sigma_{\mathrm{comb},\,i}\,
    \Sigma_{\mathrm{pred},\,i}}{\sum_{i} \Sigma_{\mathrm{comb},\,i}},
\end{equation}
where $\Sigma_{\mathrm{inte},\,i}$ and $\Sigma_{\mathrm{pred},\,i}$ are the pixel values from the respective maps.
The combined mass surface density from the total-power data, $\Sigma_{\mathrm{comb},\,i}$, serves as the statistical weight, physically anchoring the calibration to the true total mass distribution of the cloud rather than the filtered data.
The alignment factor is then simply:
\begin{equation}
    k = \frac{\bar{\Sigma}_{\mathrm{inte}}}{\bar{\Sigma}_{\mathrm{pred}}}.
\end{equation}
Across all 54 mock configurations, $k$ exhibits minor variations, ranging closely from $1.0$ to $1.2$.
The necessity and impact of this conversion factor are visually demonstrated in Figure~\ref{fig6:2d_hist}, which presents the global two-dimensional histogram pixel-by-pixel fidelity for the full configuration ensemble ($\sim$$4.5$ million valid pixels).
As shown in panel (a), before applying the alignment, the raw analytical predictions exhibit a clear linear correlation but suffer from systematic slope offsets.
Once the configuration-specific alignment factor $k$ is applied (panel (b)), the global pixel distribution converges sharply along the 1:1 relation at $\Sigma > 10^{-3}\,\mathrm{g\,cm^{-2}}$.
This confirms that the mass-weighted alignment successfully calibrates the absolute flux variations introduced by \textsc{casa} \texttt{tclean}, proving that our final calibrated CDD-\texttt{erf} transfer function faithfully reproduces the interferometric observation at the pixel level.
The practical implications of this fully calibrated framework are significant.
Conventional interferometric simulations require transforming images into the visibility domain, sampling the $uv$-plane, and applying nonlinear deconvolution algorithms, all of which incur substantial computational cost.
The CDD--\texttt{erf} framework, by contrast, operates entirely in the image domain and executes in seconds.
For theorists, it enables the rapid conversion of hydrodynamical simulations into realistic interferometric mock observations.
For observers, it provides a powerful, quantitative forward-modeling tool to estimate missing flux fractions and structural filtering effects during the preparation of observing proposals.
\subsection{The Physical Driver of Flux Loss: Spatial Scale, Not Surface Density}\label{results:physical_driver}
\begin{figure}[ht!]
\centering
\includegraphics[width=1\textwidth]{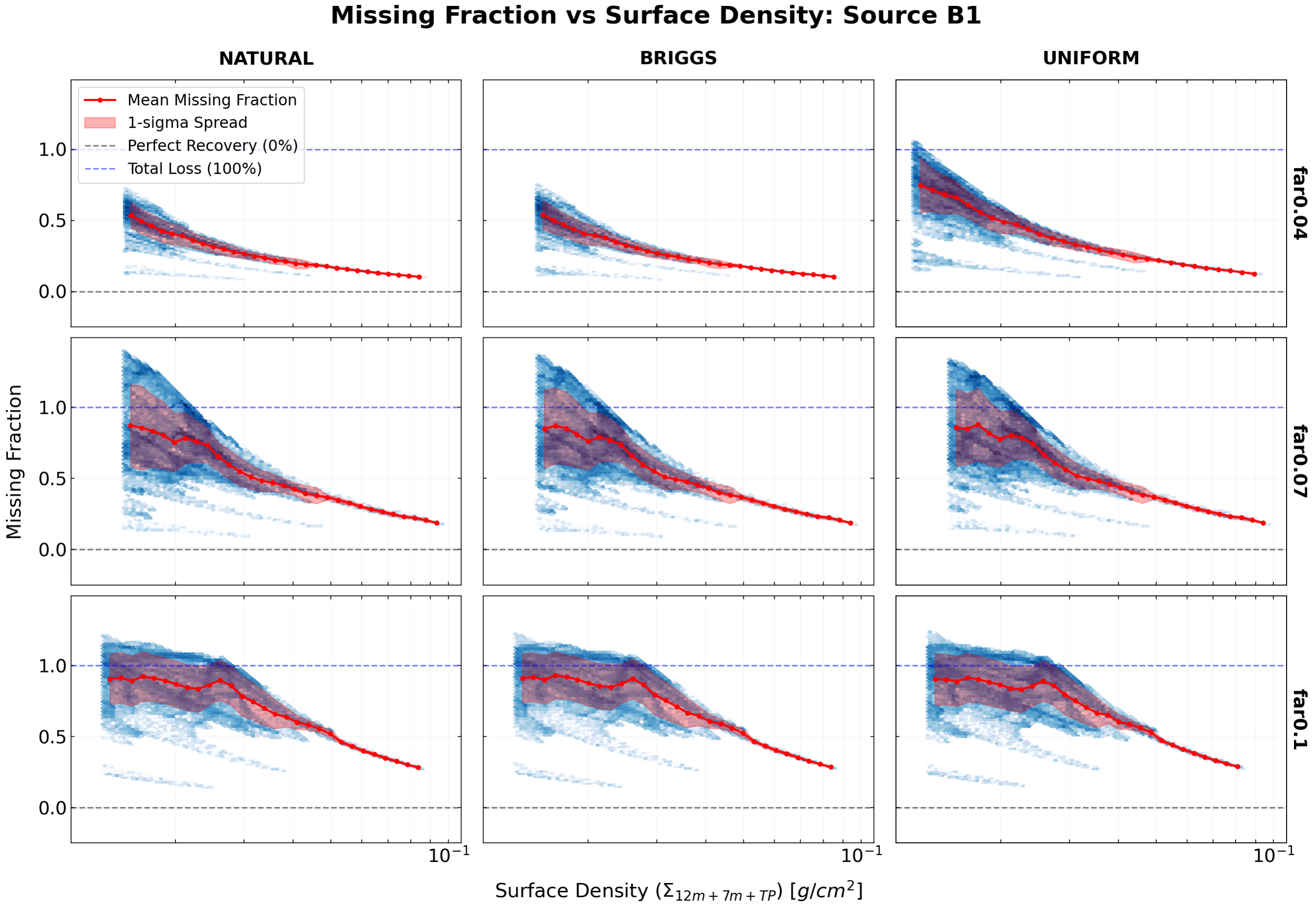}
\caption{\textbf{Missing flux fraction as a function of mass surface density for the B1
sub-region (pixels with $\Sigma_{12\,\mathrm{m} + 7\,\mathrm{m} + \mathrm{TP}}$ $> 3\sigma_{\mathrm{comb}}$).}
The red curve shows the mean missing fraction, and the shaded region denotes the $\pm 1\sigma$ dispersion.
The blue color scale indicates the pixel number density, with darker shades corresponding to higher pixel counts.
The panel layout follows Figure~\ref{fig2:Erf_Fit}.
}
\label{fig7:surface_density}
\end{figure}

In the preceding sections, we demonstrated that interferometers systematically resolve out diffuse emission.
This observational result is often interpreted as evidence that interferometric arrays are intrinsically insensitive to low-density gas.
Figure~\ref{fig7:surface_density} appears at first to support this interpretation: the mean missing flux fraction (red curve) increases monotonically as the surface density decreases, approaching zero at the highest surface densities ($\Sigma \gtrsim 10^{-1}\,\mathrm{g\,cm^{-2}}$).
However, a closer examination of the scatter in Figure~\ref{fig7:surface_density} reveals that surface density is not the controlling parameter.
In the low-density regime ($\Sigma \lesssim 10^{-2}\,\mathrm{g\,cm^{-2}}$), the missing flux fraction spans the full range from $0$ to $1$.
Data points with a missing fraction $>1$ correspond to the presence of negative bowls in the interferometric maps.
Pixels with identical low surface densities experience outcomes ranging from complete recovery to complete resolved-out.
This scatter demonstrates that surface density alone does not determine the interferometric flux loss.
The fundamental driver is spatial scale.
Because interferometers sample the sky in the Fourier $uv$-domain, due to the physical impossibility of zero-spacing baselines, they function as spatial high-pass filters.
The recovered flux fraction depends on the angular scale of the structure in which the emission resides, not on its absolute intensity or column density.
Low-density pixels that are fully recovered belong to compact, localized structures whose angular scales fall below the MRS.
Low-density pixels that suffer near-complete flux loss belong to extended background envelopes whose angular scales exceed the MRS.
This scale-filtering mechanism also accounts for the near-perfect flux recovery observed at high surface densities (right side of Figure~\ref{fig7:surface_density}).
In the interstellar medium, the highest surface densities are associated almost exclusively with gravitationally concentrated, compact structures such as dense cores and narrow filaments.
Because the physical sizes of these structures are characteristically smaller than the MRS, they are transmitted through the interferometric high-pass filter with negligible attenuation.
It follows that interferometers reliably trace the properties of compact dense cores because their angular scales are small relative to the MRS --- not simply because their column densities are high.
\section{Conclusion}
In this work, we present systematic ALMA simulations of the Perseus molecular cloud to investigate the effect of interferometric spatial filtering on the recovery of physical structures.
By combining the Constrained Diffusion Decomposition \citep{2022ApJS..259...59L} method with mass surface density analysis, we establish a quantitative framework for evaluating flux recovery across spatial scales.
We have developed the CDD-\texttt{erf} method, encapsulated in a powerful analytical master equation: $I_{\mathrm{pred}} = \sum_{l=1}^{n} [\mathrm{CDD}_l(I_{\mathrm{in}}) \times R(l)] + N_{\mathrm{rms}}$, which allows us to predict realistic interferometric images directly from physical models instantly.
Our main conclusions are as follows:
\begin{enumerate}
    \item \textbf{Sigmoid-like flux recovery curve:} The spatial filtering behavior of an interferometer acts as a scale-dependent transfer function. Provided a sufficient signal-to-noise ratio, whether a physical structure is recovered is determined by its spatial scale rather than its absolute surface density. In the CDD scale-space, this filtering is accurately modeled by a sigmoidal curve following a one-dimensional \texttt{erf}: compact structures well below the filtering limit are effectively recovered, while large-scale diffuse emission is systematically resolved out.
    \item \textbf{Characteristic transition scale:}
    The inflection point of the \texttt{erf} decay curve is intrinsically determined by the interferometer configuration.
    The characteristic transition scale can be estimated using the theoretical rule-of-thumb equation for the MRS: $\theta_{\mathrm{MRS, theoretical}} \approx 0.6\,\lambda / L_{\mathrm{min}}$ (with operational values adjusting for $uv$-coverage density, e.g., adopting the 5th percentile baseline).
    This relation provides a direct analytical mapping between the instrumental setup and the scale-space \texttt{erf} threshold ($c_{\mathrm{recover}}$).
    \item \textbf{Analytical prediction of flux loss and data alignment.} The CDD--\texttt{erf} framework directly maps the true sky brightness distribution $I_{\mathrm{in}}$ to the spatially filtered observation $I_{\mathrm{pred}}$, through interferometric flux loss as a continuous function of spatial scale.
    It enables direct alignment between different data types: by applying the derived \texttt{erf} transfer function, theoretical hydrodynamical simulations can be converted into realistic mock interferometric observations without requiring computationally expensive simulation software.
    This approach provides a quantitative bridge between theoretical models and interferometric observations.
    \item \textbf{Observational guidance for flux recovery:}
    The quantitative \texttt{erf} transfer functions across spatial scales provide practical guidance for observation planning and image reconstruction.
    For compact structures with angular scales $\lesssim 0.5\theta_{\mathrm{MRS}}$, the missing flux is limited to $\lesssim 15\%$ under suitable visibility weighting.
    As the target scale approaches the MRS, the expected flux deficit reaches approximately $50\%$.
    Structures larger than the MRS require complementary short-spacing data to avoid morphological distortion and mass underestimation.

\end{enumerate}
In summary, the CDD--\texttt{erf} framework presented here not only offers a rapid, practical tool for the analysis of multiscale astronomical data but also provides a vital quantitative bridge between theoretical hydrodynamical simulations and realistic interferometric observations.

\begin{acknowledgments}
GXL acknowledges support from NSFC grant No.12273032.
\end{acknowledgments}
\bibliography{sample7}{}

\begin{thebibliography}{}
\expandafter\ifx\csname natexlab\endcsname\relax\def\natexlab#1{#1}\fi
\providecommand{\url}[1]{\href{#1}{#1}}
\providecommand{\dodoi}[1]{doi:~\href{http://doi.org/#1}{\nolinkurl{#1}}}
\providecommand{\doeprint}[1]{\href{http://ascl.net/#1}{\nolinkurl{http://ascl.net/#1}}}
\providecommand{\doarXiv}[1]{\href{https://arxiv.org/abs/#1}{\nolinkurl{https://arxiv.org/abs/#1}}}

\bibitem[{S. {Bialy} {et~al.}(2021){Bialy}, {Zucker}, {Goodman}, {Foley}, {Alves}, {Semenov}, {Benjamin}, {Leike}, \& {En{\ss}lin}}]{2021ApJ...919L...5B}
{Bialy}, S., {Zucker}, C., {Goodman}, A., {et~al.} 2021, \bibinfo{title}{{The Per-Tau Shell: A Giant Star-forming Spherical Shell Revealed by 3D Dust Observations},} \apjl, 919, L5, \dodoi{10.3847/2041-8213/ac1f95}

\bibitem[{ {CASA Team} {et~al.}(2022){CASA Team}, {Bean}, {Bhatnagar}, {Castro}, {Donovan Meyer}, {Emonts}, {Garcia}, {Garwood}, {Golap}, {Gonzalez Villalba}, \& et~al.}]{2022PASP..134k4501C}
{CASA Team}, {Bean}, B., {Bhatnagar}, S., {et~al.} 2022, \bibinfo{title}{{CASA, the Common Astronomy Software Applications for Radio Astronomy},} \pasp, 134, 114501, \dodoi{10.1088/1538-3873/ac9642}

\bibitem[{M.~C.-Y. {Chen} {et~al.}(2024){Chen}, {Di Francesco}, {Friesen}, {Pineda}, {Caselli}, {Ginsburg}, {Kirk}, {Punanova}, \& {The GAS Collaboration}}]{2024ApJ...977..135C}
{Chen}, M. C.-Y., {Di Francesco}, J., {Friesen}, R.~K., {et~al.} 2024, \bibinfo{title}{{Filament Accretion and Fragmentation in the Perseus Molecular Cloud},} \apj, 977, 135, \dodoi{10.3847/1538-4357/ad88e8}

\bibitem[{P. Cortes {et~al.}(2025)Cortes, Carpenter, Kameno, Loomis, Vila~Vilaro, Immer, Vlahakis, Law, Stoehr, Saini, Hales, \& Kneissl}]{cortes_2025_14933753}
Cortes, P., Carpenter, J., Kameno, S., {et~al.} 2025, ALMA Cycle 12 Technical Handbook, \dodoi{10.5281/zenodo.14933753}

\bibitem[{W.~D. {Cotton}(2017){Cotton}}]{2017PASP..129i4501C}
{Cotton}, W.~D. 2017, \bibinfo{title}{{Fourier Plane Image Combination by Feathering},} \pasp, 129, 094501, \dodoi{10.1088/1538-3873/aa793f}

\bibitem[{W.~J. {Dirienzo} {et~al.}(2015){Dirienzo}, {Brogan}, {Indebetouw}, {Chandler}, {Friesen}, \& {Devine}}]{2015AJ....150..159D}
{Dirienzo}, W.~J., {Brogan}, C., {Indebetouw}, R., {et~al.} 2015, \bibinfo{title}{{Physical Conditions of the Earliest Phases of Massive Star Formation: Single-dish and Interferometric Observations of Ammonia and CCS in Infrared Dark Clouds},} \aj, 150, 159, \dodoi{10.1088/0004-6256/150/5/159}

\bibitem[{M.~L. {Enoch} {et~al.}(2006){Enoch}, {Young}, {Glenn}, {Evans}, {Golwala}, {Sargent}, {Harvey}, {Aguirre}, {Goldin}, {Haig}, \& et~al.}]{2006ApJ...638..293E}
{Enoch}, M.~L., {Young}, K.~E., {Glenn}, J., {et~al.} 2006, \bibinfo{title}{{Bolocam Survey for 1.1 mm Dust Continuum Emission in the c2d Legacy Clouds. I. Perseus},} \apj, 638, 293, \dodoi{10.1086/498678}

\bibitem[{S. {Faridani} {et~al.}(2018){Faridani}, {Bigiel}, {Fl{\"o}er}, {Kerp}, \& {Stanimirovi{\'c}}}]{2018AN....339...87F}
{Faridani}, S., {Bigiel}, F., {Fl{\"o}er}, L., {Kerp}, J., \& {Stanimirovi{\'c}}, S. 2018, \bibinfo{title}{{A new approach for short-spacing correction of radio interferometric datasets},} Astronomische Nachrichten, 339, 87, \dodoi{10.1002/asna.201713381}

\bibitem[{J.~A. {H{\"o}gbom}(1974){H{\"o}gbom}}]{1974A&AS...15..417H}
{H{\"o}gbom}, J.~A. 1974, \bibinfo{title}{{Aperture Synthesis with a Non-Regular Distribution of Interferometer Baselines},} \aaps, 15, 417

\bibitem[{J. {Kauffmann} {et~al.}(2017){Kauffmann}, {Pillai}, {Zhang}, {Menten}, {Goldsmith}, {Lu}, \& {Guzm{\'a}n}}]{2017A&A...603A..89K}
{Kauffmann}, J., {Pillai}, T., {Zhang}, Q., {et~al.} 2017, \bibinfo{title}{{The Galactic Center Molecular Cloud Survey. I. A steep linewidth-size relation and suppression of star formation},} \aap, 603, A89, \dodoi{10.1051/0004-6361/201628088}

\bibitem[{J. {Koda} {et~al.}(2019){Koda}, {Teuben}, {Sawada}, {Plunkett}, \& {Fomalont}}]{2019PASP..131e4505K}
{Koda}, J., {Teuben}, P., {Sawada}, T., {Plunkett}, A., \& {Fomalont}, E. 2019, \bibinfo{title}{{Total Power Map to Visibilities (TP2VIS): Joint Deconvolution of ALMA 12m, 7m, and Total Power Array Data},} \pasp, 131, 054505, \dodoi{10.1088/1538-3873/ab047e}

\bibitem[{R.~B. {Larson}(1981){Larson}}]{1981MNRAS.194..809L}
{Larson}, R.~B. 1981, \bibinfo{title}{{Turbulence and star formation in molecular clouds.},} \mnras, 194, 809, \dodoi{10.1093/mnras/194.4.809}

\bibitem[{G.-X. {Li}(2022){Li}}]{2022ApJS..259...59L}
{Li}, G.-X. 2022, \bibinfo{title}{{Multiscale Decomposition of Astronomical Maps: A Constrained Diffusion Method},} \apjs, 259, 59, \dodoi{10.3847/1538-4365/ac4bc4}

\bibitem[{G.-X. {Li} {et~al.}(2016){Li}, {Burkert}, {Megeath}, \& {Wyrowski}}]{2016arXiv160305720L}
{Li}, G.-X., {Burkert}, A., {Megeath}, T., \& {Wyrowski}, F. 2016, \bibinfo{title}{{Gravitational acceleration and edge effects in molecular clouds},} arXiv e-prints, arXiv:1603.05720, \dodoi{10.48550/arXiv.1603.05720}

\bibitem[{G.-X. {Li} {et~al.}(2015){Li}, {Wyrowski}, {Menten}, {Megeath}, \& {Shi}}]{2015A&A...578A..97L}
{Li}, G.-X., {Wyrowski}, F., {Menten}, K., {Megeath}, T., \& {Shi}, X. 2015, \bibinfo{title}{{G-virial: Gravity-based structure analysis of molecular clouds},} \aap, 578, A97, \dodoi{10.1051/0004-6361/201424030}

\bibitem[{G.-X. {Li} \& J.-X. {Zhou}(2022){Li} \& {Zhou}}]{2022MNRAS.514L..16L}
{Li}, G.-X., \& {Zhou}, J.-X. 2022, \bibinfo{title}{{Density exponent analysis: gravity-driven steepening of the density profiles of star-forming regions},} \mnras, 514, L16, \dodoi{10.1093/mnrasl/slac049}

\bibitem[{T. {Lindeberg}(1994{\natexlab{a}}){Lindeberg}}]{1994JApSt..21..225L}
{Lindeberg}, T. 1994{\natexlab{a}}, \bibinfo{title}{{Scale-space theory: a basic tool for analyzing structures at different scales},} Journal of Applied Statistics, 21, 225, \dodoi{10.1080/757582976}

\bibitem[{T. {Lindeberg}(1994{\natexlab{b}}){Lindeberg}}]{1994sstc.book.....L}
{Lindeberg}, T. 1994{\natexlab{b}}, {Scale-space theory in computer vision}

\bibitem[{M. {Lombardi} {et~al.}(2010){Lombardi}, {Lada}, \& {Alves}}]{2010A&A...512A..67L}
{Lombardi}, M., {Lada}, C.~J., \& {Alves}, J. 2010, \bibinfo{title}{{2MASS wide field extinction maps. III. The Taurus, Perseus, and California cloud complexes},} \aap, 512, A67, \dodoi{10.1051/0004-6361/200912670}

\bibitem[{U. {Mahmut} {et~al.}(2024){Mahmut}, {Esimbek}, {Baan}, {Tang}, {Zhou}, {Li}, {He}, {Tursun}, {Li}, {Komesh}, \& et~al.}]{2024MNRAS.528..577M}
{Mahmut}, U., {Esimbek}, J., {Baan}, W., {et~al.} 2024, \bibinfo{title}{{Formaldehyde observations of the Perseus Molecular Cloud},} \mnras, 528, 577, \dodoi{10.1093/mnras/stad3959}

\bibitem[{G.~N. {Ortiz-Le{\'o}n} {et~al.}(2018){Ortiz-Le{\'o}n}, {Loinard}, {Dzib}, {Galli}, {Kounkel}, {Mioduszewski}, {Rodr{\'\i}guez}, {Torres}, {Hartmann}, {Boden}, \& et~al.}]{2018ApJ...865...73O}
{Ortiz-Le{\'o}n}, G.~N., {Loinard}, L., {Dzib}, S.~A., {et~al.} 2018, \bibinfo{title}{{The Gould{\textquoteright}s Belt Distances Survey (GOBELINS). V. Distances and Kinematics of the Perseus Molecular Cloud},} \apj, 865, 73, \dodoi{10.3847/1538-4357/aada49}

\bibitem[{A. {Plunkett} {et~al.}(2023){Plunkett}, {Hacar}, {Moser-Fischer}, {Petry}, {Teuben}, {Pingel}, {Kunneriath}, {Takagi}, {Miyamoto}, {Moravec}, {Suri}, {Hess}, {Hoffman}, \& {Mason}}]{2023PASP..135c4501P}
{Plunkett}, A., {Hacar}, A., {Moser-Fischer}, L., {et~al.} 2023, \bibinfo{title}{{Data Combination: Interferometry and Single-dish Imaging in Radio Astronomy},} \pasp, 135, 034501, \dodoi{10.1088/1538-3873/acb9bd}

\bibitem[{J.~F. {Radcliffe} {et~al.}(2024){Radcliffe}, {Beswick}, {Thomson}, {Njeri}, \& {Muxlow}}]{2024MNRAS.527..942R}
{Radcliffe}, J.~F., {Beswick}, R.~J., {Thomson}, A.~P., {Njeri}, A., \& {Muxlow}, T.~W.~B. 2024, \bibinfo{title}{{Revisiting a flux recovery systematic error arising from common deconvolution methods used in aperture-synthesis imaging},} \mnras, 527, 942, \dodoi{10.1093/mnras/stad2694}

\bibitem[{U. {Rau} {et~al.}(2019){Rau}, {Naik}, \& {Braun}}]{2019AJ....158....3R}
{Rau}, U., {Naik}, N., \& {Braun}, T. 2019, \bibinfo{title}{{A Joint Deconvolution Algorithm to Combine Single-dish and Interferometer Data for Wideband Multiterm and Mosaic Imaging},} \aj, 158, 3, \dodoi{10.3847/1538-3881/ab1aa7}

\bibitem[{N.~A. {Ridge} {et~al.}(2006){Ridge}, {Di Francesco}, {Kirk}, {Li}, {Goodman}, {Alves}, {Arce}, {Borkin}, {Caselli}, {Foster}, \& et~al.}]{2006AJ....131.2921R}
{Ridge}, N.~A., {Di Francesco}, J., {Kirk}, H., {et~al.} 2006, \bibinfo{title}{{The COMPLETE Survey of Star-Forming Regions: Phase I Data},} \aj, 131, 2921, \dodoi{10.1086/503704}

\bibitem[{F.~R. {Schwab}(1984){Schwab}}]{1984AJ.....89.1076S}
{Schwab}, F.~R. 1984, \bibinfo{title}{{Relaxing the isoplanatism assumption in self-calibration; applications to low-frequency radio interferometry},} \aj, 89, 1076, \dodoi{10.1086/113605}

\bibitem[{S. {Stanimirovic}(2002){Stanimirovic}}]{2002ASPC..278..375S}
{Stanimirovic}, S. 2002, in Astronomical Society of the Pacific Conference Series, Vol. 278, Single-Dish Radio Astronomy: Techniques and Applications, ed. S.~{Stanimirovic}, D.~{Altschuler}, P.~{Goldsmith}, \& C.~{Salter}, 375--396, \dodoi{10.48550/arXiv.astro-ph/0205329}

\bibitem[{J.-L. {Starck} \& F. {Murtagh}(2006){Starck} \& {Murtagh}}]{2006aida.book.....S}
{Starck}, J.-L., \& {Murtagh}, F. 2006, {Astronomical Image and Data Analysis}, \dodoi{10.1007/978-3-540-33025-7}

\bibitem[{M. {Tahani} {et~al.}(2022){Tahani}, {Lupypciw}, {Glover}, {Plume}, {West}, {Kothes}, {Inutsuka}, {Lee}, {Robishaw}, {Knee}, \& et~al.}]{2022A&A...660A..97T}
{Tahani}, M., {Lupypciw}, W., {Glover}, J., {et~al.} 2022, \bibinfo{title}{{3D magnetic-field morphology of the Perseus molecular cloud},} \aap, 660, A97, \dodoi{10.1051/0004-6361/202141170}

\bibitem[{A.~R. {Thompson} {et~al.}(2017){Thompson}, {Moran}, \& {Swenson}}]{2017isra.book.....T}
{Thompson}, A.~R., {Moran}, J.~M., \& {Swenson}, Jr., G.~W. 2017, {Interferometry and Synthesis in Radio Astronomy, 3rd Edition}, \dodoi{10.1007/978-3-319-44431-4}

\bibitem[{J.~J. {Tobin} {et~al.}(2016){Tobin}, {Looney}, {Li}, {Chandler}, {Dunham}, {Segura-Cox}, {Sadavoy}, {Melis}, {Harris}, {Kratter}, \& et~al.}]{2016ApJ...818...73T}
{Tobin}, J.~J., {Looney}, L.~W., {Li}, Z.-Y., {et~al.} 2016, \bibinfo{title}{{The VLA Nascent Disk and Multiplicity Survey of Perseus Protostars (VANDAM). II. Multiplicity of Protostars in the Perseus Molecular Cloud},} \apj, 818, 73, \dodoi{10.3847/0004-637X/818/1/73}

\bibitem[{D. {Wright} {et~al.}(2025){Wright}, {Ad{\'a}mek}, \& {Armour}}]{2025ASPC..541..155W}
{Wright}, D., {Ad{\'a}mek}, K., \& {Armour}, W. 2025, in Astronomical Society of the Pacific Conference Series, Vol. 541, Astronomical Data Analysis Software and Systems XXXIII, ed. A.~{Jacques}, R.~{Seaman}, N.~{Gandilo}, \& T.~{Linder}, 155

\bibitem[{E. {Zari} {et~al.}(2016){Zari}, {Lombardi}, {Alves}, {Lada}, \& {Bouy}}]{2016A&A...587A.106Z}
{Zari}, E., {Lombardi}, M., {Alves}, J., {Lada}, C.~J., \& {Bouy}, H. 2016, \bibinfo{title}{{Herschel-Planck dust optical depth and column density maps. II. Perseus},} \aap, 587, A106, \dodoi{10.1051/0004-6361/201526597}

\bibitem[{C. {Zhang} {et~al.}(2022){Zhang}, {Zhang}, {Li}, \& {Li}}]{2022RAA....22e5012Z}
{Zhang}, C., {Zhang}, G.-Y., {Li}, J.-Z., \& {Li}, X.-M. 2022, \bibinfo{title}{{Herschel Investigation of Cores and Filamentary Structures in the Perseus Molecular Cloud},} Research in Astronomy and Astrophysics, 22, 055012, \dodoi{10.1088/1674-4527/ac5bc7}

\bibitem[{C. {Zucker} {et~al.}(2021){Zucker}, {Goodman}, {Alves}, {Bialy}, {Koch}, {Speagle}, {Foley}, {Finkbeiner}, {Leike}, {En{\ss}lin}, \& et~al.}]{2021ApJ...919...35Z}
{Zucker}, C., {Goodman}, A., {Alves}, J., {et~al.} 2021, \bibinfo{title}{{On the Three-dimensional Structure of Local Molecular Clouds},} \apj, 919, 35, \dodoi{10.3847/1538-4357/ac1f96}

\end{thebibliography}
\bibliographystyle{aasjournalv7}
\clearpage
\appendix
\vspace{-0.3in}
\setcounter{figure}{0}
\renewcommand{\thefigure}{A\arabic{figure}}
\section{Input Sky Model and Sub-regions}
\begin{figure}[H]
\centering
\includegraphics[width=0.89\textwidth]{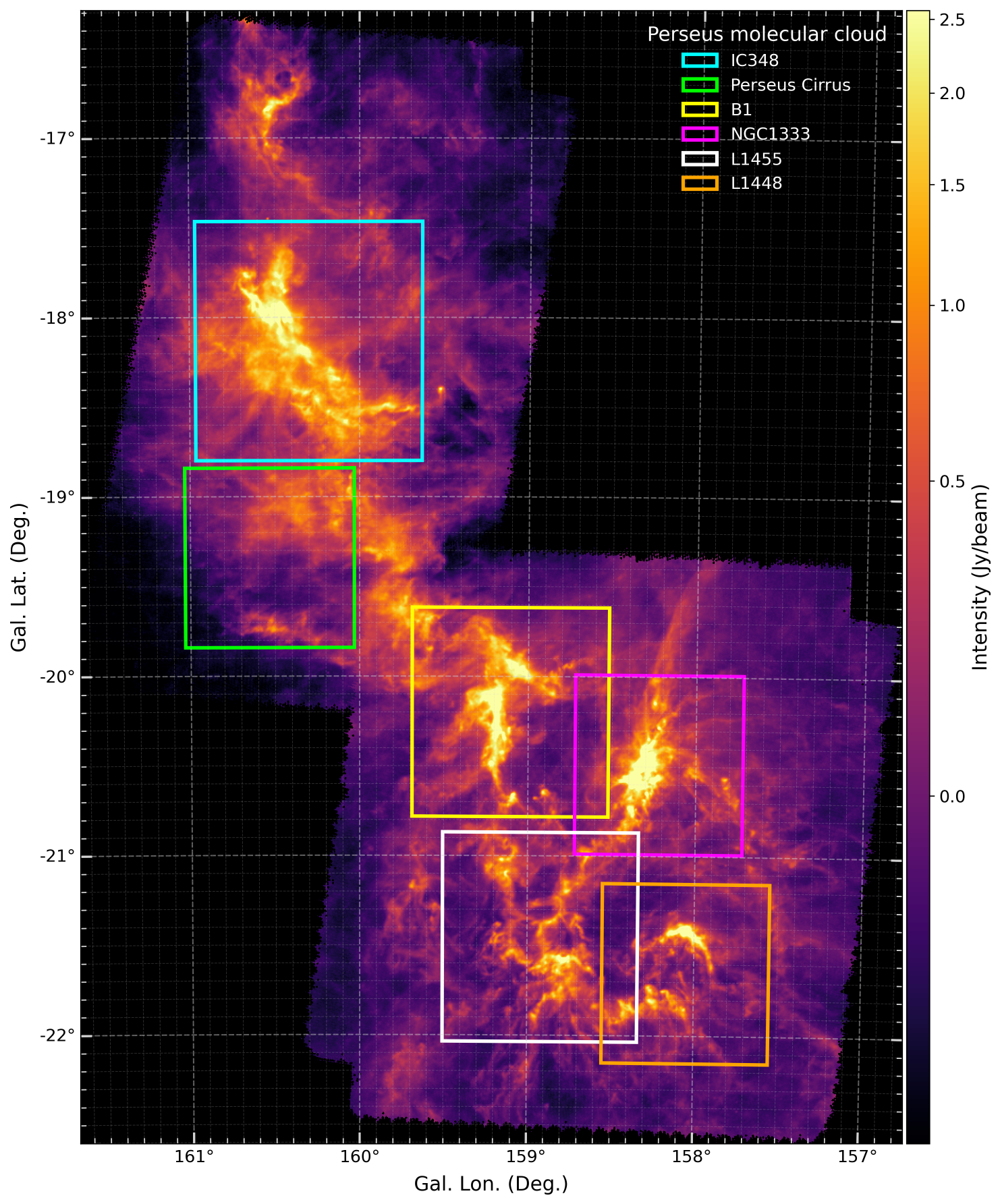}
\caption{\textbf{\textit{Herschel}/SPIRE $500\,\mu\mathrm{m}$ mosaic map of the Perseus molecular cloud.}
This image serves as the input sky model for our \textsc{casa} interferometric simulations.
Colored boxes indicate the six sub-regions selected for independent multiscale analysis: IC~348 (cyan), Perseus Cirrus (green), B1 (yellow), NGC~1333 (magenta), L1455 (white), and L1448 (orange).
The color bar shows the emission intensity in units of $\mathrm{Jy\,beam^{-1}}$.
}
\label{fig:appendix_map}
\end{figure}
\section{Morphological reconstruction}
\setcounter{figure}{0}
\renewcommand{\thefigure}{B\arabic{figure}}
\begin{figure}[ht!]
\centering
\includegraphics[width=0.95\textwidth]{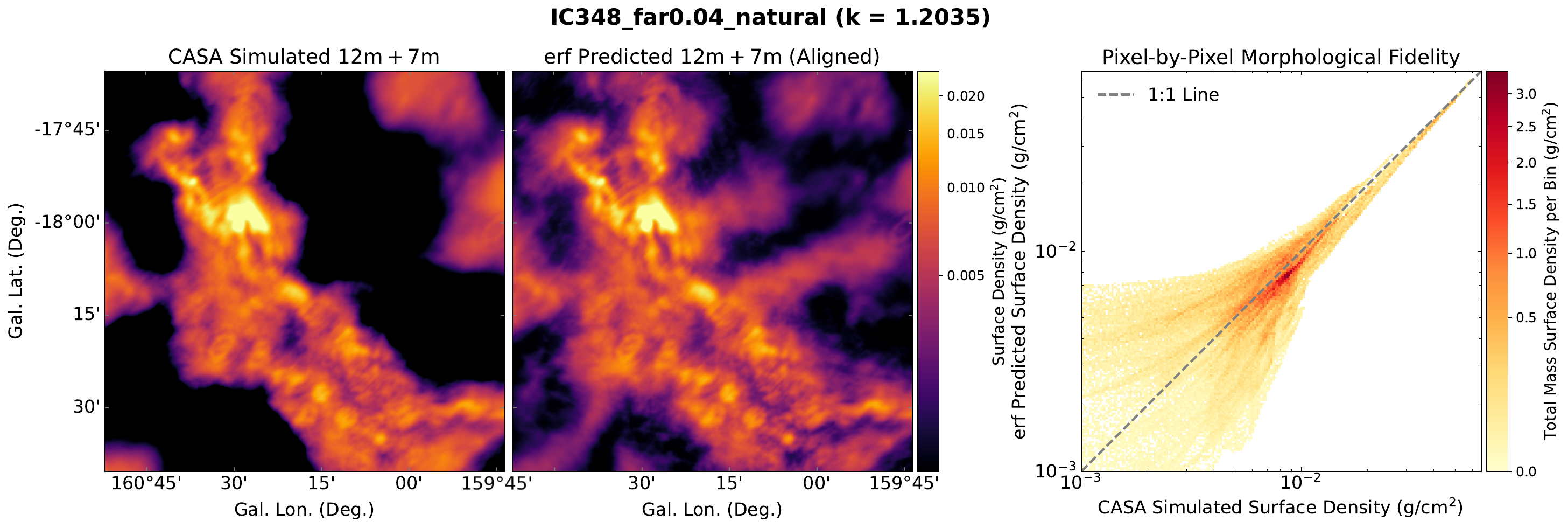}\vspace{0.3cm}
\includegraphics[width=0.95\textwidth]{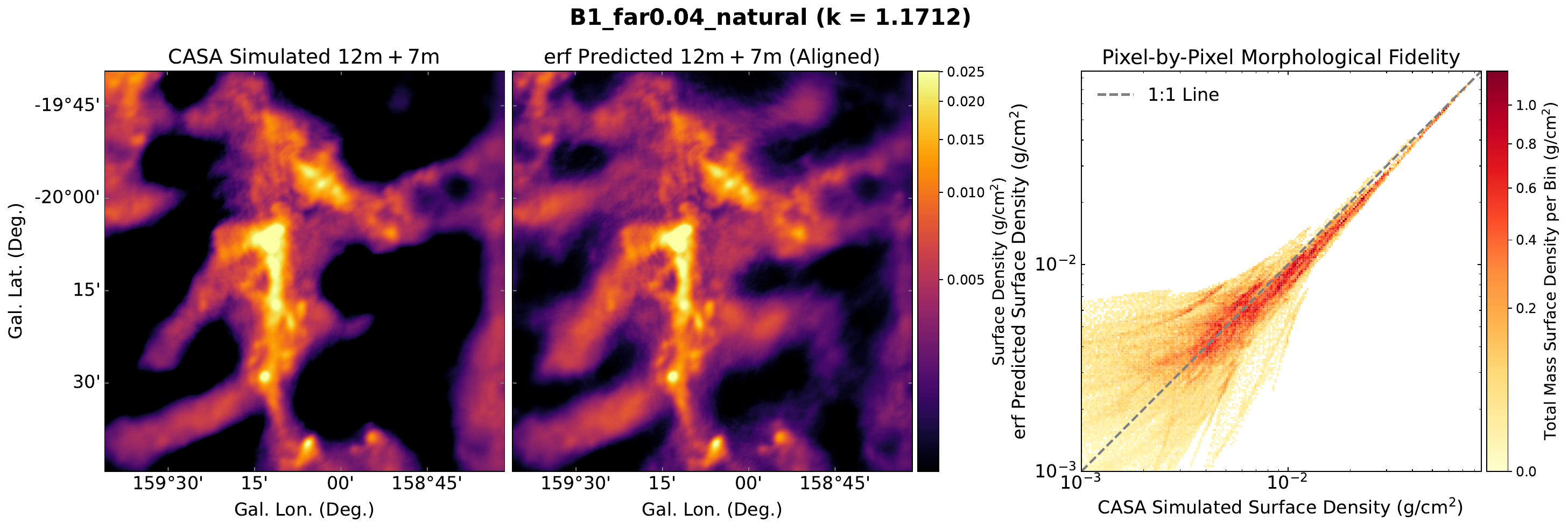}\vspace{0.3cm}
\includegraphics[width=0.95\textwidth]{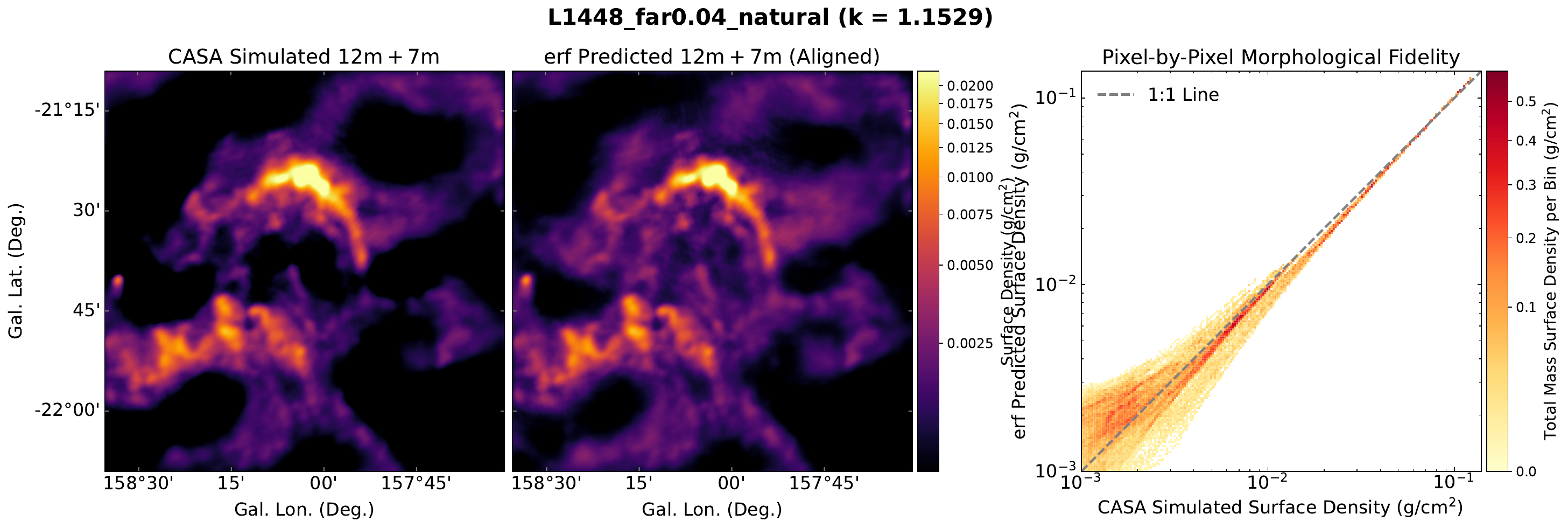}
\caption{\textbf{Morphological reconstruction of the remaining five sub-regions (B1, IC348, L1448, L1455, and Perseus Cirrus) under the \texttt{far0.04} configuration with natural weighting.}
For each sub-region, the left and center panels show the
\textsc{casa}-simulated $12\,\mathrm{m}+7\,\mathrm{m}$ mass surface density map and the aligned CDD--\texttt{erf}-predicted map, respectively, with the mass-weighted mean alignment factor $k$ indicated in each panel header.
The right panel shows the corresponding pixel-by-pixel two-dimensional histogram.
All axis conventions and pixel selection criteria follow Figure~\ref{fig5:example_ngc1333}. (Figure continued on the following page.)}
\label{fig:appendix_other_sources}
\end{figure}
\clearpage
\begin{figure}[ht!]
\addtocounter{figure}{-1}
\centering
\includegraphics[width=0.95\textwidth]{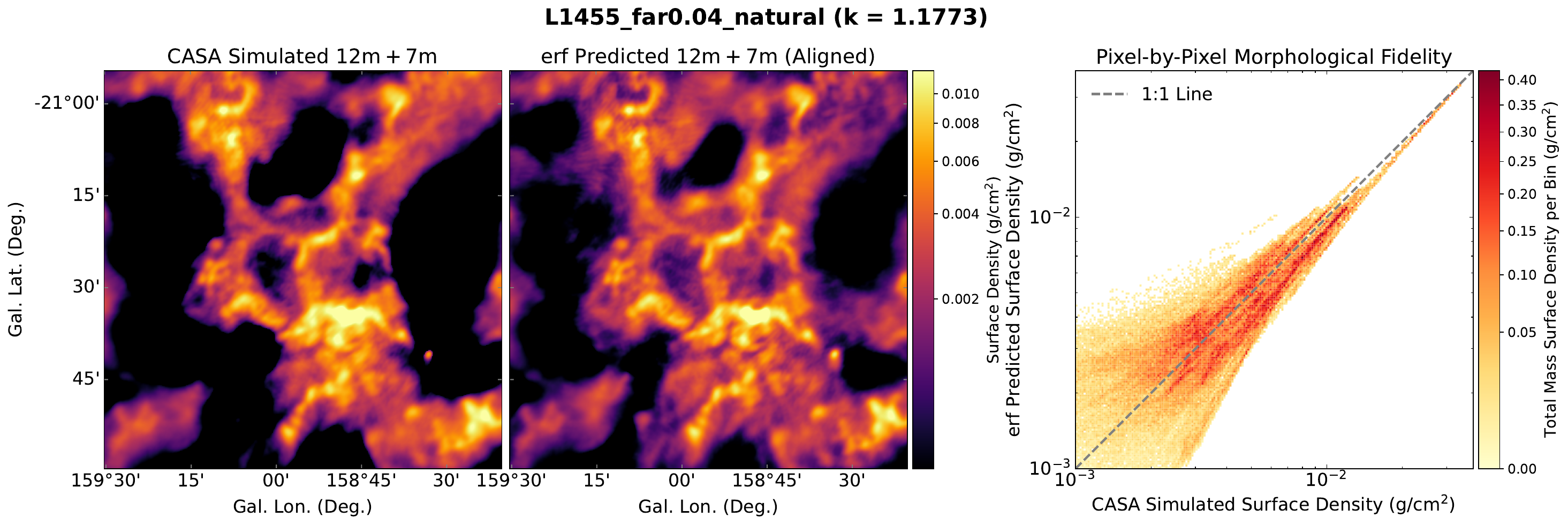}\vspace{0.3cm}
\includegraphics[width=0.95\textwidth]{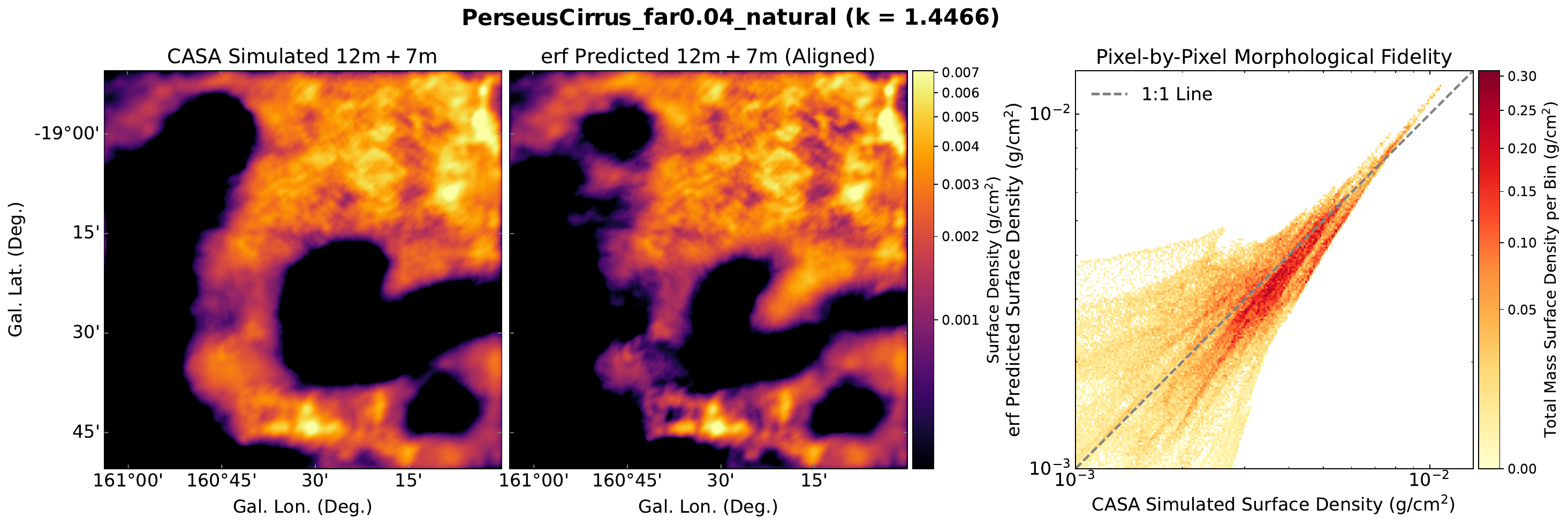}
\caption{\textit{Continued.}}
\end{figure}
\section{Python Implementation of the CDD-erf Predictor}
\label{app:python_code}
The entire analytical forward-modeling workflow, encapsulated by Equation~(\ref{eq:master_equation}), can be executed in a few lines of Python code.
Before running the script, install the CDD package from \url{https://github.com/gxli/Constrained-Diffusion-Decomposition}.
We provide a minimal, executable script below demonstrating how to predict an interferometric image directly from an input sky model.
To facilitate community use, this Python script (CDD erf Predictor.py) is provided as a supplementary downloadable file associated with this article.
\begin{lstlisting}[language=Python, basicstyle=\ttfamily\small, keywordstyle=\bfseries, stringstyle=\ttfamily, commentstyle=\itshape]
#The CDD-erf Predictor
import numpy as np
from scipy.special import erf
from astropy.io import fits
import constrained_diffusion as cdd
import matplotlib.pyplot as plt
from astropy.visualization import simple_norm

def get_erf_transfer_function(scale_index, c_recover, B, w):
    """ Equation erf: R(l) = (B/2) * [1 - erf((l - c_recover) / w)] """
    return B * 0.5 * (1 - erf((scale_index - c_recover) / w))

def predict_interferometric_image(input_fits, output_fits, erf_params, sigma_rms=0.0):
    """
    equation: I_pred = Sum [ CDD_l(I_in) * R(l) ] + N_rms
    """
    # 1. Load the input sky model (I_in)
    header = fits.getheader(input_fits)
    I_in = fits.getdata(input_fits).squeeze()
    I_in = np.nan_to_num(I_in, nan=0.0)

    # 2. Perform CDD decomposition: CDD_l(I_in)
    print("Running CDD decomposition...")
    components, residual = cdd.constrained_diffusion_decomposition(I_in)

    # Initialize the predicted image (I_pred)
    I_pred = np.zeros_like(I_in)

    # 3. Apply the transfer function R(l) and sum the components
    c_recover, B, w = erf_params['c_recover'], erf_params['B'], erf_params['w']

    for i, comp in enumerate(components):
        l = i + 1  # scale index (l)
        R_l = get_erf_transfer_function(l, c_recover, B, w)
        I_pred += comp * R_l  # I_pred += CDD_l(I_in) * R(l)

    # 4. Add thermal noise (N_rms, optional)
    if sigma_rms > 0.0:
        print(f"Adding thermal noise (N_rms) with sigma = {sigma_rms}...")
        N_rms = np.random.normal(loc=0.0, scale=sigma_rms, size=I_pred.shape)
        I_pred += N_rms

    # 5. Save the spatially filtered interferometric image
    fits.writeto(output_fits, I_pred, header, overwrite=True)
    print(f"Prediction saved to {output_fits}")
    return I_pred


# Example Usage:
if __name__ == "__main__":
    # ALMA C-4 array configuration parameters
    params = {'c_recover': 7.5, 'B': 0.90, 'w': 1.50}
    user_sigma_rms = 1e-4
    # Execute prediction
    I_pred = predict_interferometric_image('input.fits', 'predicted_ALMA.fits', params, sigma_rms=user_sigma_rms)
    # Quick Visualization
    norm = simple_norm(I_pred, stretch='asinh', min_percent=1.0, max_percent=99.5)
    plt.imshow(I_pred, origin='lower', cmap='inferno', norm = norm)
    plt.show()
\end{lstlisting}

\vspace{12pt}
\noindent\textbf{ORCID:}\\
Dan Miao: \url{https://orcid.org/0000-0001-9798-9852}\\
Guang-Xing Li: \url{https://orcid.org/0000-0003-3144-1952}

\end{document}